\documentclass[a4paper]{panl}
\usepackage{cite}
\usepackage{wrapfig}
\usepackage{graphicx}
\usepackage{amssymb}
\usepackage{amsfonts}
\usepackage{amsmath}
\usepackage{longtable}
\usepackage{rotating}
\usepackage{lscape}
\usepackage{epsfig}
\usepackage{multirow}

\def\CF{\mathrm{C_F}}
\def\CA{\mathrm{C_A}}
\def\Nc{\mathrm{N_c}}
\def\W{\mathcal{W}}
\def\WI{\overline{\mathcal{W}}}
\def\A{\mathcal{A}}
\def\bA{\bar{\mathcal{A}}}
\def\bB{\bar{\mathcal{B}}}
\def\bC{\bar{\mathcal{C}}}
\def\B{\mathcal{B}}
\def\C{\mathcal{C}}
\def\D{\mathcal{D}}
\def\E{\mathcal{E}}
\def\F{\mathcal{F}}
\def\I{\mathcal{I}}
\def\G{\mathcal{G}}
\def\X{\mathcal{X}}
\def\inn{\mathrm{in}}
\def\out{\mathrm{out}}
\def\NG{\mathrm{NG}}
\renewcommand{\d}{\mathrm{d}}
\newcommand{\as}{\alpha_s}
\newcommand{\Li}{\,\mathrm{Li}}
\def\R{{\scriptscriptstyle\mathrm{R}}}
\def\V{{\scriptscriptstyle\mathrm{V}}}
\renewcommand{\S}{\mathcal{S}}

\originalTeX

\begin{document}
% Journal sections (see http://pkp.jinr.ru/index.php/PEPAN_LETTERS/about/editorialPolicies#focusAndScope)
\issuearea{Physics of Elementary Particles and Atomic Nuclei. Theory}
% or in Russian
%\issuearea{ФИЗИКА ЭЛЕМЕНТАРНЫХ ЧАСТИЦ И АТОМНОГО ЯДРА. ТЕОРИЯ}

\title{Eikonal amplitudes and non-global logarithms\\from the BMS equation}
\maketitle
\authors{
H. Benslama$^{a,}$\footnote{E-mail: hana.benslama@univ-batna.dz},
Y. Delenda$^{a,}$\footnote{E-mail: yazid.delenda@univ-batna.dz (Corresponding author.)},
K. Khelifa-Kerfa$^{b,}$\footnote{E-mail: kamel.kkhelifa@iu.edu.sa},
A. M. Ibrahim$^{a,}$\footnote{E-mail: abdelaziz.alrufai@iu.edu.sa}}
\from{$^{a}$ Laboratoire de Physique des Rayonnements et de leurs Int\'{e}ractions avec la Mati\`{e}re\\
D\'{e}partement de Physique, Facult\'{e} des Sciences de la Mati\`{e}re\\
Universit\'{e} de Batna-1, Batna 05000, Algeria}
\vspace{-3mm}
\from{$^{b}$ Department of Physics, Faculty of Science\\
Islamic University of Madinah, Madinah 42351, Saudi Arabia}

\begin{abstract}
% Russian translation of the abstract
%Аннотация на русском языке. \
%\vspace{0.2cm}
The Banfi-Marchesini-Smye (BMS) equation accounts for resummation of non-global logarithms to all orders in perturbation theory in the large-$\Nc$ approximation. We show that the squared amplitudes for the emission of soft energy-ordered gluons are correctly embedded in this equation, and explicitly verify that they coincide with those derived in our previous work in the large-$\Nc$ limit up to sixth order in the strong coupling. We perform analytical calculations for the non-global logarithms up to fourth order for the specific hemisphere mass distribution in $e^+e^-$ collisions, thus confirming our previous semi-numerical results. We show that the solution to the BMS equation may be cast into a product of an infinite number of exponentials each of which resums a class of Feynman diagrams that manifest a symmetry pattern, and explicitly carry out the computation of the first of these exponentials.
\end{abstract}
\vspace*{6pt}

%\noindent
%KEYWORDS: QCD; resummation; non-global logarithms; eikonal amplitudes


\section*{Introduction}	

Achieving precision in the calculation of QCD observables at particle colliders is usually hindered by many perturbative and non-perturbative issues. The resummation of large logarithms is one such issue. Uncanceled virtual emissions above some scale $Q_0$ up to a renormalization scale $\mu_\mathrm{R}$ contribute to the integrated distribution of the observable in question when real emissions above $Q_0$ are cut, leading to logarithms of the form $\as^n\ln^m(\mu_\mathrm{R}/Q_0)$, with $m\leq 2n$ for observables sensitive to soft and/or collinear emissions (e.g., jet mass), while $m\leq n$ for observables sensitive to soft wide-angle emissions alone (e.g., energy flow into gaps between jets). In this paper we consider the former case of double logarithmic enhancements, i.e., up to $(\as\,L^2)^n$, with $L=\ln(\mu_\mathrm{R}/Q_0)$.

In the leading logarithmic (LL) accuracy all double logarithms of the form $\as^nL^{n+1}$ in the exponent of the distribution are resummed, and in the next-to-leading logarithmic (NLL) accuracy all single logarithms $\as^nL^n$ are additionally resummed. The resummation of these large logarithms is particularly quite simple for observables that are inclusive over emissions in the entire angular phase space since typically a limited number of gluon emissions needs to be considered in order to reproduce the all-orders behavior of the distribution of the said observables.

NLL accuracy in many QCD observables that are referred to as non-global, i.e., those sensitive to emissions in restricted regions of the angular phase space, has proven difficult to reach due to their cumbersome resummation. Emissions outside the ``unmeasured'' region which themselves subsequently emit a single softest gluon back into the ``measured'' region lead to a tower of large single logarithms $\as^n L^n$ \cite{Dasgupta:2001sh, Dasgupta:2002bw}. The treatment of these non-global logarithms (NGLs) relies on multiple gluon branching that becomes increasingly complicated at higher orders in perturbation theory. Their resummation, contrary to that of global observables mentioned above, cannot be performed by a consideration of a fixed number of gluon emissions since an iterative pattern could not thus far be spotted.

The topic of resummation of NGLs has seen a lot of interest in the literature. NGLs were first spotted and numerically resummed in the large-$\Nc$ limit ($\Nc$ is the number of quark colors) by Dasgupta and Salam in Refs. \cite{Dasgupta:2001sh,Dasgupta:2002bw}. Banfi, Marchesini, and Smye derived a non-linear integro-differential equation, the BMS equation \cite{Banfi:2002hw}, whose solution resums NGLs at large $\Nc$ for emissions off a given dipole. A numerical solution to this equation was provided in the same reference for away-from-jets energy flow observable. An analytic solution to the BMS equation was also achieved by Schwartz and Zhu in Ref. \cite{Schwartz:2014wha} by means of a perturbative expansion in the strong coupling up to fifth order. In Ref. \cite{Larkoski:2016zzc}, the convergence of the ``dressed gluon expansion'' \cite{Caron-Huot:2015bja} of the BMS equation was studied and the results of this technique were compared to the numerical series-solution of the BMS equation up to 12 orders. An analogous integro-differential equation that accounts for NGLs to all orders at finite $\Nc$ was also proposed by Weigert in Ref. \cite{Weigert:2003mm}, and was numerically solved in Ref. \cite{Hatta:2013iba} in the context of away-from-jets energy flow, and in Ref. \cite{Hagiwara:2015bia} for the hemisphere mass distribution in $e^+e^-\to q\bar{q}$ events. Additionally, an evolution algorithm that deals with the resummation of NGLs at finite $\Nc$ to all orders was developed in Refs. \cite{Martinez:2018ffw,Forshaw:2019ver}. In Ref. \cite{Khelifa-Kerfa:2015mma}, we evaluated NGLs at finite $\Nc$ up to fifth order in the coupling by computing eikonal amplitudes of soft gluon emissions (as in Ref. \cite{Delenda:2015tbo}) and using phase-space considerations. NGLs have also been considered in the context of groomed multi-prong jet shapes in Ref. \cite{Neill:2018yet}. Progress has also been made towards resumming NGLs beyond single logarithmic accuracy as discussed in Refs. \cite{Caron-Huot:2015bja, Becher:2015hka}.

In this work we consider the integrated hemisphere mass distribution in the process $e^+e^-\to q\bar{q}$, and propose a solution to the BMS equation for NGLs in the form of an exponential. The exponent is written as a series in the strong coupling (or equivalently the ``evolution'' parameter), and we show how it may be resummed by illustrating the partial resummation of the two-loops term in the exponent. The results we find are consistent with those obtained by Schwartz and Zhu in Ref. \cite{Schwartz:2014wha}, which were written as an expansion in the coupling in the actual distribution and not in the exponent. The benefit of writing the series in the exponent is to avoid obvious interference terms that pop up in the perturbative expansion of the distribution. The proposed solution paves the way for a clearer picture for the all-orders resummation as it subtracts away any such interference. Additionally, the present work serves as a stringent test on the validity of both the BMS equation and the squared eikonal amplitudes derived and used in Refs. \cite{Khelifa-Kerfa:2015mma, Delenda:2015tbo}, by showing how the said amplitudes at large $\Nc$ are correctly reproduced by the BMS equation up to six loops. Recall that the eikonal amplitudes in Ref. \cite{Delenda:2015tbo} were computed at finite $\Nc$.

This paper is organized as follows. In section 2 we introduce our notation and define the hemisphere mass observable. In section 3 we write down the BMS equation and propose its solution as an exponential of a series in the coupling. We explicitly treat each order in the exponent up to sixth order, hence we extract the squared eikonal amplitudes from the BMS equation and compare with previous results in the literature, in section 4. In section 5 we perform necessary integrations in order to obtain the coefficients of the NGLs in the exponent of the hemisphere mass distribution up to four loops. We show that the expansion of our result agrees with that found in Refs. \cite{Schwartz:2014wha,Khelifa-Kerfa:2015mma}. We then show how the resummation of the terms in the exponent may be achieved by considering the resummation of the two-loops contribution. Finally, in section 6, we draw our conclusions.

\section*{Kinematics, notation, and observable }

To illustrate the resummation of NGLs and extract eikonal amplitudes from the BMS equation we consider a simple observable, namely the hemisphere mass in di-jet events in $e^+e^-$ collisions, where the hard scattering is accompanied by strongly-ordered soft gluon emissions $\omega_n\ll\cdots\ll\omega_2\ll\omega_1\ll Q$, with $Q$ the hard scale and $\omega_i$ the energy of the $i^\mathrm{th}$ emission. The four-momenta of the outgoing quark ($a$), anti-quark $(b)$, and gluons $(i)$ are given by
\begin{subequations}
\begin{align}
p_a&=\frac{Q}{2}\left(1,0,0,1 \right),\\
p_b&=\frac{Q}{2}\left(1,0,0,-1 \right),\\
k_i&=\omega_i\left(1,\sin\theta_i\cos\phi_i,\sin\theta_i\sin\phi_i,\cos\theta_i\right),
\end{align}
\end{subequations}
where recoil effects are negligible at single-logarithmic accuracy. Here $\theta_i$ and $\phi_i$ are the polar and azimuthal angles of the $i^\mathrm{th}$ emission.

To compare our findings of eikonal amplitudes to those presented in Ref. \cite{Delenda:2015tbo} we follow the same notation introduced therein. We define the dipole \emph{antenna} functions, which carry the angular structure of the squared matrix elements, as follows
\begin{subequations}
\begin{align}
w_{ab}^i&=\omega_i^2\,\frac{p_a\cdot p_b}{(p_a\cdot k_i)\,(k_i\cdot p_b)}\,,\label{eq:omega}\\
\A_{ab}^{ij}&=w_{ab}^i\left(w_{ai}^j+w_{ib}^j-w_{ab}^j\right),\\
\B_{ab}^{ijk}&=w_{ab}^i\left(\A_{ai}^{jk}+\A_{ib}^{jk}-\A_{ab}^{jk}\right),\\
\C_{ab}^{ijk\ell}&=w_{ab}^i\left(\B_{ai}^{jk\ell}+\B_{ib}^{jk\ell}-\B_{ab}^{jk\ell}\right),\\
\D_{ab}^{ijk\ell m}&=w_{ab}^i\left(\C_{ai}^{jk\ell m}+\C_{ib}^{jk\ell m}-\C_{ab}^{jk\ell m}\right),\\
\E_{ab}^{ijk\ell mn}&=w_{ab}^i\left(\D_{ai}^{jk\ell mn}+\D_{ib}^{jk\ell mn}-\D_{ab}^{jk\ell mn}\right).
\end{align}
\end{subequations}
We additionally define the reduced antenna functions $\bA_{ab}^{ij}=\A_{ab}^{ij}/w_{ab}^i$, and similarly for the other functions. The basic antenna function $w_{ab}^i$ may be cast into
\begin{equation}
\begin{split}
w_{ab}^i&=\frac{(ab)}{(ai)\,(ib)}\,,\\
(ij)&=1-\cos\theta_{ij}=1-c_i\,c_j-s_i\,s_j\cos\phi_{ij}\,,
\end{split}
\end{equation}
where for compactness we define $c_i\equiv\cos\theta_i$, $s_i\equiv\sin\theta_i$, and $\phi_{ij}=\phi_i-\phi_j$.

The observable we are interested in is the normalized (squared) invariant mass of the right hemisphere $\mathcal{H}_\mathrm{R}$ (in the direction of the quark $(a)$)
\begin{align}
\rho&=\frac{1}{Q^2}\left(p_a+\sum_{i\in\mathcal{H}_\mathrm{\mathrm{R}}}k_i\right)^2\notag\\
&=\frac{1}{Q^2}\sum_{i\in\mathcal{H}_\mathrm{R}}2\,k_i\cdot p_a=\sum_{i\in\mathcal{H}_\mathrm{R}}x_i\left(1-\cos\theta_i\right),
\end{align}
where the sum runs over all gluon emissions inside the right hemisphere $\mathcal{H}_\mathrm{R}$, and $x_i=\omega_i/Q$ is the energy fraction of the $i^{\mathrm{th}}$ emission.

\section*{The BMS equation and its solution}

The hemisphere mass distribution is sensitive to soft and/or collinear gluon emissions leading to large logarithms $L=\ln(1/\rho)$ in the integrated distribution $\sigma(\rho)$ of this observable. At leading order in color (also known as the large-$\Nc$ approximation), this distribution satisfies a non-linear integro-differential equation known as the BMS equation \cite{Banfi:2002hw}
\begin{equation}\label{eq:BMS}
\frac{\partial\G_{ab}(t)}{\partial t}=\Nc\int\frac{\d\Omega_k}{4\pi}\,w_{ab}^k\left(\Theta^\out_k\,\G_{ak}(t)\,\G_{kb}(t)-\G_{ab}(t)\right),
\end{equation}
where the hemisphere mass distribution is just $\sigma(\rho)=\mathcal{G}_{ab}(t)$. The evolution parameter $t$ is related to $\rho$ by
\begin{equation}
t=\frac{1}{\pi}\int_\rho^1\frac{\d x}{x}\,\as(Q\,x)=-\frac{1}{2\pi\beta_0}\,\ln\left(1-2\as\beta_0L\right),
\end{equation}
with $\beta_0$ the one-loop coefficient of the QCD $\beta$ function and the second equality holds at one loop. In Eq. \eqref{eq:BMS} $\d\Omega_k=\d c_k\,\d\phi_k$ is the differential solid angle of the emission $k$, and the step function $\Theta^\out_k$ restricts this emission to be outside the measured hemisphere $\mathcal{H}_\mathrm{R}$, thus $\Theta^{\mathrm{out}}_k=\Theta(-c_k)$. We additionally have $\Theta^\inn_k=1-\Theta^\out_k=\Theta(c_k)$.

The quantity $\G_{ab}(t)$ is generally interpreted as the probability that a given dipole $(ab)$ whose directions are determined by solid angles $\Omega_a$ and $\Omega_b$ emits radiation resulting in a hemisphere mass less than $\rho$. We shall confine ourselves in the present work to the specific dipole $(ab)=(q\bar{q})$ fixed by $c_a=+1$ and $c_b=-1$, or equivalently $\theta_a=0$ and $\theta_b=\pi$. The solution to the BMS equation is unique with the initial condition $\mathcal{G}_{ab}(t=0)=1$. Typical values of $t$ for phenomenological studies go up to $t\sim 0.3$.

The first term $\Theta^\out_k\,\G_{ak}(t)\,\G_{kb}(t)$ in Eq. \eqref{eq:BMS} represents real-emission contributions while the subtracted term $-\G_{ab}(t)$ represents the virtual corrections, and they are both (when integrated) separately divergent when the emission is collinear to one of the hard dipole legs. This collinear singularity is cancelled in the sum of the two terms. To avoid these divergences we rewrite the BMS equation in an alternative way
\begin{align}\label{eq:BMS2}
\frac{\partial\G_{ab}(t)}{\partial t}&=\Nc\int\frac{\d\Omega_k}{4\pi}\,w_{ab}^k\,\Theta^\out_k\left(\G_{ak}(t)\,\G_{kb}(t)-\G_{ab}(t)\right)-\notag\\
&-\Nc\int\frac{\d\Omega_k}{4\pi}\,w_{ab}^k\,\Theta^\inn_k\,\G_{ab}(t)\,.
\end{align}
Eq. \eqref{eq:BMS2} as it stands is still ill-defined, since there is another collinear singularity in the second term associated with emissions parallel to the quark direction. This issue was also raised in Ref. \cite{Hagiwara:2015bia}. In fact, the BMS equation was originally written for away-from-jets energy flow observables which are free from contributions of such emissions. Observables of this kind are sensitive only to soft wide-angle emissions and thus only have single logarithms. On the contrary, the hemisphere mass distribution has both single and double logarithms. This explains the origin of the divergence in Eq. \eqref{eq:BMS2}. In order for the BMS equation to be valid in our case a kinematical cutoff needs to be applied to the second term in Eq. \eqref{eq:BMS2}, namely a collinear cutoff $1-c_k>\rho$. This cutoff naturally arises when using the ``measurement-operator"  procedure to account for real and virtual contributions to the hemisphere mass distribution as explained in Ref. \cite{Khelifa-Kerfa:2015mma}.

This collinear singularity, however, only affects the resummation of double logarithms, which typically results in a Sudakov form factor, and does not enter the resummation of NGLs. To see this we recall that NGLs result from emissions near the boundary between the two hemispheres which is away from the quark and anti-quark directions \cite{Dasgupta:2002bw}. Said differently, collinear emissions to the quark or anti-quark do not contribute to NGLs. Since we are interested only in the resummation of NGLs in this paper, and since the double logarithmic (Sudakov) form factor has been well treated before (see Ref. \cite{Dasgupta:2002dc} and the Review \cite{Dasgupta:2003iq}), we can therefore discard this divergence in the current paper.

\subsection*{Exponential solution}

Based on the observation made in Ref. \cite{Khelifa-Kerfa:2015mma} about the possible exponentiation of NGLs, and given that the derivative of the function $\G_{ab}(t)$ in the BMS equation reproduces a phase-space integral of somewhat the same function, it is natural to propose an exponential solution with a series in the exponent of the form
\begin{equation}\label{eq:proposed}
\G_{ab}(t)=\exp\left(\sum_{n=1}^\infty\frac{1}{n!}\,\S_{ab}^{(n)}\,\mathrm{N_c^{\it n}}\,t^n\right),
\end{equation}
where $\S_{ab}^{(n)}$ are fixed coefficients. This proposed solution satisfies the initial condition mentioned above at $t=0$ and includes the primary-emission Sudakov form factor (the term $n=1$ in the exponent). This exponential form avoids the unnecessary dealing with ``interference'' terms discussed in Ref. \cite{Khelifa-Kerfa:2015mma}, which can be reproduced here by expanding the exponential. The BMS equation does in fact admit this solution since, as we shall see, all the coefficients $\S_{ab}^{(n)}$ in the exponent are fully determined recursively down to the Sudakov term $\S_{ab}^{(1)}$, which too is fully determined.

Although the BMS equation is valid only at leading color, one can write the series in the exponent in terms of the color factors $\CF=(\mathrm{N_c^2}-1)/(2\,\Nc)$ and $\CA=\Nc$ by merely replacing $\mathrm{N_c^{\it n}}\to 2\,\CF\,\mathrm{C_A^{{\it n}-1}}$, where at large $\Nc$ we have $2\,\CF=\CA= \Nc$. This fact helps us \emph{partially} restore the full color structure of the resummed NGLs at finite $\Nc$. Further finite-$\Nc$ corrections totally not accounted for by the BMS equation, and which contribute a factor $\sigma_{\mathrm{corr}}$ (first appearing at $\mathcal{O}(\as^4)$ \cite{Khelifa-Kerfa:2015mma}) to the fully resummed distribution, are also required, such that
\begin{equation}
\begin{split}
\sigma_{\mathrm{full\,}\Nc}(\rho)&=\sigma_{\mathrm{large\,}\Nc}(\rho)\times\sigma_{\mathrm{corr}}(\rho)\,,\\
\sigma_{\mathrm{corr}}(\rho)&=1+\mathcal{O}(\as^4)\,.
\end{split}
\end{equation}
The first term in $\sigma_{\mathrm{corr}}(\rho)$ was computed in Ref. \cite{Khelifa-Kerfa:2015mma} to be
\begin{equation}
\frac{1}{4!}\,\bar{\alpha}_s^4\,L^4\,\CF\,\mathrm{C_A^3}\left(\frac{1}{2}-\frac{\CF}{\CA}\right)\zeta_4\,,
\end{equation}
with $\bar{\alpha}_s=\alpha_s/\pi$ and $\zeta$ is the Riemann zeta function.

\subsection*{Iteration of the series coefficients}

We now show how the coefficients in the exponent of the exponential solution may be found iteratively. Taking the derivative of the solution \eqref{eq:proposed} with respect to $t$ and renaming the summation index we obtain
\begin{equation}
\frac{\partial\G_{ab}(t)}{\partial t}=\left(\sum_{n=0}^\infty\frac{1}{n!}\,\S_{ab}^{(n+1)}\,\mathrm{N_c^{{\it n}+1}}\,t^n\right)\times\G_{ab}(t)\,,
\end{equation}
where we note that the sum here starts at $n=0$. Substituting into the BMS equation and dividing both sides by $\Nc\,\G_{ab}(t)$ we obtain
\begin{align}
\sum_{n=0}^\infty\frac{1}{n!}\,\mathcal{S}_{ab}^{(n+1)}\,(\Nc\,t)^n=&-\int\frac{\d\Omega_k}{4\pi}\,\Theta^\inn_k\,w_{ab}^k+\int\frac{\d\Omega_k}{4\pi}\,\Theta^\out_k\,w_{ab}^k\times\notag\\
&\times\left(\exp\left[\sum_{n=1}^\infty\frac{1}{n!}\left(\S_{ak}^{(n)}+\S_{kb}^{(n)}-\S_{ab}^{(n)}\right)(\Nc\,t)^n\right]-1\right).
\end{align}
In order to extract the coefficients $\S_{ab}^{(n)}$ at any order $n$ it suffices to equate coefficients of $(\Nc\,t)^n$ from both sides of this equation. At zeroth order (equating coefficients of $(\Nc\,t)^0$) we have
\begin{equation}\label{eq:Sab1}
\S_{ab}^{(1)}=-\int\frac{\d\Omega_1}{4\pi}\,\Theta^\inn_1\,w_{ab}^1\,,
\end{equation}
where we have set $k=1$ to represent the first emission. This is the coefficient in the exponent of the Sudakov form factor. As stated earlier, this term is divergent and can be regulated by placing a collinear cutoff on the polar integration. However, this divergence is irrelevant for the calculation of NGLs and thus will not be considered further.

At higher orders we may write
\begin{align}
\sum_{n=1}^\infty\frac{1}{n!}\,\S_{ab}^{(n+1)}\,(\Nc\,t)^n=&\int\frac{\d\Omega_1}{4\pi}\,\Theta^\out_1\,w_{ab}^1\times\notag\\
&\times\left(\exp\left[\sum_{n=1}^\infty\frac{1}{n!}\left(\S_{a1}^{(n)}+\S_{1b}^{(n)}-\S_{ab}^{(n)}\right)(\Nc\,t)^n\right]-1\right).
\end{align}
Denoting $\X_n=\S_{a1}^{(n)}+ \S_{1b}^{(n)}-\S_{ab}^{(n)}$, and using the fact that
\begin{align}
&\exp\left(\sum_{n=1}^\infty\frac{(\Nc\,t)^n}{n!}\,\X_n\right)-1=\frac{(\Nc\,t)^1}{1!}\,\X_1+\frac{(\Nc\,t)^2}{2!}\left(\X_1^2+\X_2\right)+\notag\\
&+\frac{(\Nc\,t)^3}{3!}\left(\X_1^3+3\,\X_1\,\X_2+\X_3\right)+\notag\\
&+\frac{(\Nc\,t)^4}{4!}\left(\X_1^4+6\,\X_1^2\,\X_2+3\,\X_2^2+4\,\X_1\,\X_3+\X_4\right)+\notag\\
&+\frac{(\Nc\,t)^5}{5!}\left(\X_1^5+10\X_1^3\X_2+15\X_1\X_2^2+10\X_1^2\X_3+10\X_2\X_3+5\X_1\X_4+\X_5\right)+\notag\\
&+\frac{(\Nc\,t)^6}{6!}\left(\X_1^6+15\,\X_1^4\,\X_2+45\,\X_1^2\,\X_2^2+15\,\X_2^3+20\,\X_1^3\,\X_3+60\,\X_1\,\X_2\,\X_3+\right.\notag\\
&\qquad\qquad\quad\left.+10\,\X_3^2+15\,\X_1^2\,\X_4+15\,\X_2\,\X_4+6\,\X_1\,\X_5+\X_6\right)+\mathcal{O}(t^7)\notag\\
&\equiv\sum_{n=1}^\infty\frac{1}{n!}\,\F^{(n)}_{ab}(k_1)\,(\Nc\,t)^n\,,
\end{align}
we are able to compute all the coefficients $\S_{ab}^{(n)}$ recursively
\begin{align}
\S_{ab}^{(n+1)}=\int\frac{\d\Omega_1}{4\pi}\,\Theta_1^\out\,w_{ab}^1\,\F_{ab}^{(n)}(k_1)\,,\qquad n\geq1\,,
\end{align}
where we note that all $\F_{ab}^{(n)}(k_1)$ are written as combinations of the functions $\X_m$ with $m \leq n$. This means that the $(n+1)^{\mathrm{th}}$-order coefficient $\S_{ab}^{(n+1)}$ is written in terms of lower-order terms recursively. In what follows below we illustrate the evaluation of these coefficients up to sixth order. Going to higher orders is in principle possible though cumbersome.

\subsection*{Results up to six loops}

The leading order at which NGLs first appear is the two-loops order, that is $\mathcal{O}(\as^2)$. At this order we have
\begin{align}\label{eq:S2}
\S_{ab}^{(2)}&=\int\frac{\d\Omega_1}{4\pi}\,\Theta^\out_1\,w_{ab}^1\,\F_{ab}^{(1)}(k_1)\notag\\
&= \int\frac{\d\Omega_1}{4\pi}\,\Theta^\out_1\,w_{ab}^1\,\X_1\notag\\
&= \int\frac{\d\Omega_1}{4\pi}\,\Theta^\out_1\,w_{ab}^{1}\left(\S_{a1}^{(1)}+\S_{1b}^{(1)}-\S_{ab}^{(1)}\right)\notag\\
&=-\int\frac{\d\Omega_1}{4\pi}\frac{\d\Omega_2}{4\pi}\,\Theta^\out_1\,\Theta^\inn_2\,w_{ab}^{1}\left(w_{a1}^2+w_{1b}^2-w_{ab}^2\right)\notag\\
&=-\int\frac{\d\Omega_{12}}{(4\pi)^2}\,\Theta^\out_1\,\Theta^\inn_2\,\A_{ab}^{12}\,,
\end{align}
where we substituted the expressions for $\S_{ij}^{(1)}$ from the relation \eqref{eq:Sab1}, and used the shorthand notation $\d\Omega_{12\hdots m}\equiv\d\Omega_1\,\d\Omega_2\,\hdots\,\d\Omega_m$.

At three loops we find
\begin{align}\label{eq:S3}
\S_{ab}^{(3)}=&\int\frac{\d\Omega_1}{4\pi}\,\Theta^\out_1\,w_{ab}^1\left[\left(\S_{a1}^{(1)}+\S_{1b}^{(1)}-\S_{ab}^{(1)}\right)^2+\S_{a1}^{(2)}+\S_{1b}^{(2)}-\S_{ab}^{(2)}\right]\notag\\
=&\int\frac{\d\Omega_{123}}{(4\pi)^3}\,\Theta^\out_1\,\Theta^\inn_2\,\Theta^\inn_3\,\A_{ab}^{12}\,\bA_{ab}^{13}-\int\frac{\d\Omega_{123}}{(4\pi)^3}\,\Theta^\out_1\,\Theta^\out_2\,\Theta^\inn_3\,\B_{ab}^{123}\,.
\end{align}
Similarly, at four loops we have the result
\begin{align}\label{eq:S4}
\S_{ab}^{(4)}=&\int\frac{\d\Omega_1}{4\pi}\,\Theta^\out_1\,w_{ab}^1\bigg[\left(\S_{a1}^{(1)}+\S_{1b}^{(1)}-\S_{ab}^{(1)}\right)^3+\S_{a1}^{(3)}+\S_{1b}^{(3)}-\S_{ab}^{(3)}+\notag\\
&+3\left(\S_{a1}^{(1)}+\S_{1b}^{(1)}-\S_{ab}^{(1)}\right)\left(\S_{a1}^{(2)}+\S_{1b}^{(2)}-\S_{ab}^{(2)}\right)\bigg]\notag\\
=&\,-\int\frac{\d\Omega_{1234}}{(4\pi)^4}\,\Theta^\out_1\,\Theta^\inn_2\,\Theta^\inn_3\,\Theta^\inn_4\,\A_{ab}^{12}\,\bA_{ab}^{13}\,\bA_{ab}^{14}+\notag\\
  &+3\int\frac{\d\Omega_{1234}}{(4\pi)^4}\,\Theta^\out_1\,\Theta^\inn_2\,\Theta^\out_3\,\Theta^\inn_4\,\A_{ab}^{12}\,\bB_{ab}^{134}+\notag\\
   &+\int\frac{\d\Omega_{1234}}{(4\pi)^4}\,\Theta^\out_1\,\Theta^\out_2\,\Theta^\inn_3\,\Theta^\inn_4\,\mathfrak{A}_{ab}^{1234}-\notag\\
   &-\int\frac{\d\Omega_{1234}}{(4\pi)^4}\,\Theta^\out_1\,\Theta^\out_2\,\Theta^\out_3\,\Theta^\inn_4\,\C_{ab}^{1234}\,,
\end{align}
where, inline with the notation used in Ref. \cite{Delenda:2015tbo}, we introduced
\begin{equation}
\mathfrak{A}_{ab}^{ijk\ell}=w_{ab}^i\left(\A_{ai}^{jk}\bA_{ai}^{j\ell}+\A_{ib}^{jk}\bA_{ib}^{j\ell}-\A_{ab}^{jk}\bA_{ab}^{j\ell}\right).
\end{equation}

Furthermore, to present the five-loops coefficient we introduce, as in Ref. \cite{Delenda:2015tbo}
\begin{subequations}
\begin{align}
\mathbb{A}_{ab}^{ijk\ell m}&=w_{ab}^i\left(\A_{ai}^{jk}\bA_{ai}^{{j\ell}}\bA_{ai}^{jm}+\A_{ib}^{jk}\bA_{ib}^{j\ell}\bA_{ib}^{jm}-\A_{ab}^{jk}\bA_{ab}^{j\ell}\bA_{ab}^{jm}\right),\\
\tilde{\mathfrak{A}}_{ab}^{ijk\ell m}&=w_{ab}^i\left(\A_{ai}^{jk}\bB_{ai}^{j\ell m}+\A_{ib}^{jk}\bB_{ib}^{j\ell m}-\A_{ab}^{jk}\bB_{ab}^{j\ell m}\right),\\
\mathfrak{B}^{ijk\ell m}_{ab} &= w_{ab}^i \left(\mathfrak{A}_{ai}^{jk\ell m} +\mathfrak{A}_{ib}^{jk\ell m} -\mathfrak{A}_{ab}^{jk\ell m} \right).
\end{align}
\end{subequations}
Then, the five-loops coefficient reads
\begin{align}
\S_{ab}^{(5)}=&\int\frac{\d\Omega_1}{4\pi}\,w_{ab}^1\,\Theta^\out_1\bigg[\left(\S_{a1}^{(1)}+\S_{1b}^{(1)}-\S_{ab}^{(1)}\right)^4+\S_{a1}^{(4)}+\S_{1b}^{(4)}-\S_{ab}^{(4)}+\notag\\
&+6\left(\S_{a1}^{(1)}+\S_{1b}^{(1)}-\S_{ab}^{(1)}\right)^2\left(\S_{a1}^{(2)}+\S_{1b}^{(2)}-\S_{ab}^{(2)}\right)+\notag\\
&+3\left(\S_{a1}^{(2)}+\S_{1b}^{(2)}-\S_{ab}^{(2)}\right)^2+\notag\\
&+4\left(\S_{a1}^{(1)}+\S_{1b}^{(1)}-\S_{ab}^{(1)}\right)\left(\S_{a1}^{(3)}+\S_{1b}^{(3)}-\S_{ab}^{(3)}\right)\bigg]\,.
\end{align}
Substituting in terms of the antenna functions we find at this order
\begin{align}\label{eq:S5}
\S_{ab}^{(5)}=&
   \int\frac{\d\Omega_{12345}}{(4\pi)^5}\,\Theta^\out_1\,\Theta^\inn_2\,\Theta^\inn_3\,\Theta^\inn_4\,\Theta^\inn_5\,\A_{ab}^{12}\,\bA_{ab}^{13}\,\bA_{ab}^{14}\,\bA_{ab}^{15}-\notag\\
&-6\int\frac{\d\Omega_{12345}}{(4\pi)^5}\,\Theta^\out_1\,\Theta^\out_2\,\Theta^\inn_3\,\Theta^\inn_4\,\Theta^\inn_5\,\B_{ab}^{123}\,\bA_{ab}^{14}\,\bA_{ab}^{15}+\notag\\
&+3\int\frac{\d\Omega_{12345}}{(4\pi)^5}\,\Theta^\out_1\,\Theta^\out_2\,\Theta^\inn_3\,\Theta^\out_4\,\Theta^\inn_5\,\B_{ab}^{123}\,\bB_{ab}^{145}-\notag\\
&-4\int\frac{\d\Omega_{12345}}{(4\pi)^5}\,\Theta^\out_1\,\Theta^\out_2\,\Theta^\inn_3\,\Theta^\inn_4\,\Theta^\inn_5\,\A_{ab}^{15}\,\bar{\mathfrak{A}}_{ab}^{1234}+\notag\\
&+4\int\frac{\d\Omega_{12345}}{(4\pi)^5}\,\Theta^\out_1\,\Theta^\out_2\,\Theta^\out_3\,\Theta^\inn_4\,\Theta^\inn_5\,\A_{ab}^{15}\,\bC_{ab}^{1234}-\notag\\
&- \int\frac{\d\Omega_{12345}}{(4\pi)^5}\,\Theta^\out_1\,\Theta^\out_2\,\Theta^\inn_3\,\Theta^\inn_4\,\Theta^\inn_5\,\mathbb{A}_{ab}^{12345}+\notag\\
&+3\int\frac{\d\Omega_{12345}}{(4\pi)^5}\,\Theta^\out_1\,\Theta^\out_2\,\Theta^\inn_3\,\Theta^\out_4\,\Theta^\inn_5\,\tilde{\mathfrak{A}}_{ab}^{12345}+\notag\\
&+ \int\frac{\d\Omega_{12345}}{(4\pi)^5}\,\Theta^\out_1\,\Theta^\out_2\,\Theta^\out_3\,\Theta^\inn_4\,\Theta^\inn_5\,\mathfrak{B}^{12345}_{ab}-\notag\\
&- \int\frac{\d\Omega_{12345}}{(4\pi)^5}\,\Theta^\out_1\,\Theta^\out_2\,\Theta^\out_3\,\Theta^\out_4\,\Theta^\inn_5\,\D_{ab}^{12345}\,.
\end{align}

For the presentation of the six-loops result we shall need the following definitions
\begin{subequations}
\begin{align}
\mathbb{B}_{ab}^{ijk\ell mn} &= w_{ab}^i\left(\mathbb{A}_{ai}^{jk\ell mn}+\mathbb{A}_{ib}^{jk\ell mn}-\mathbb{A}_{ab}^{jk\ell mn}\right),\\
\tilde{\mathfrak{B}}^{ijk\ell mn}_{ab} &= w_{ab}^i \left(\tilde{\mathfrak{A}}_{ai}^{jk\ell mn} +\tilde{\mathfrak{A}}_{ib}^{jk\ell mn} -\tilde{\mathfrak{A}}_{ab}^{jk\ell mn} \right),\\
\mathfrak{C}^{ijk\ell mn}_{ab} &= w_{ab}^i \left(\mathfrak{B}_{ai}^{jk\ell mn} +\mathfrak{B}_{ib}^{jk\ell mn} -\tilde{\mathfrak{B}}_{ab}^{jk\ell mn} \right).
\end{align}
\end{subequations}
At sixth order we have
\begin{align}
\S_{ab}^{(6)}=\int\frac{\d\Omega_{1}}{4\pi}\,w_{ab}^1\,\Theta^\out_1&\bigg[\left(\S_{a1}^{(1)}+\S_{1b}^{(1)}-\S_{ab}^{(1)}\right)^5+\S_{a1}^{(5)}+\S_{1b}^{(5)}-\S_{ab}^{(5)}+\notag\\
&+10\left(\S_{a1}^{(1)}+\S_{1b}^{(1)}-\S_{ab}^{(1)}\right)^3\left(\S_{a1}^{(2)}+\S_{1b}^{(2)}-\S_{ab}^{(2)}\right)+\notag\\
&+15\left(\S_{a1}^{(1)}+\S_{1b}^{(1)}-\S_{ab}^{(1)}\right)\left(\S_{a1}^{(2)}+\S_{1b}^{(2)}-\S_{ab}^{(2)}\right)^2+\notag\\
&+10\left(\S_{a1}^{(1)}+\S_{1b}^{(1)}-\S_{ab}^{(1)}\right)^2 \left(\S_{a1}^{(3)}+\S_{1b}^{(3)}-\S_{ab}^{(3)}\right)+\notag\\
&+10\left(\S_{a1}^{(2)}+\S_{1b}^{(2)}-\S_{ab}^{(2)}\right)\left(\S_{a1}^{(3)}+\S_{1b}^{(3)}-\S_{ab}^{(3)}\right)+\notag\\
&+5\left(\S_{a1}^{(1)}+\S_{1b}^{(1)}-\S_{ab}^{(1)}\right)\left(\S_{a1}^{(4)}+\S_{1b}^{(4)}-\S_{ab}^{(4)}\right)\bigg]\,.
\end{align}
Explicitly written we have
\begin{align}
\S_{ab}^{(6)}=&
 -  \int\frac{\d\Omega_{1\dots6}}{(4\pi)^6}\,\Theta^\out_1\,\Theta^\out_2\,\Theta^\out_3\,\Theta^\out_4\,\Theta^\out_5\,\Theta^\inn_6\,\E_{ab}^{123456}-\notag\\
&-  \int\frac{\d\Omega_{1\dots6}}{(4\pi)^6}\,\Theta^\out_1\,\Theta^\inn_2\,\Theta^\inn_3\,\Theta^\inn_4\,\Theta^\inn_5\,\Theta^\inn_6\,\A_{ab}^{12}\,\bA_{ab}^{13}\,\bA_{ab}^{14}\,\bA_{ab}^{15}\,\bA_{ab}^{16}+\notag\\
&+10\int\frac{\d\Omega_{1\dots6}}{(4\pi)^6}\,\Theta^\out_1\,\Theta^\inn_2\,\Theta^\inn_3\,\Theta^\inn_4\,\Theta^\out_5\,\Theta^\inn_6\,\A_{ab}^{12}\,\bA_{ab}^{13}\,\bA_{ab}^{14}\,\bB_{ab}^{156}-\notag\\
&-15\int\frac{\d\Omega_{1\dots6}}{(4\pi)^6}\,\Theta^\out_1\,\Theta^\inn_2\,\Theta^\out_3\,\Theta^\inn_4\,\Theta^\out_5\,\Theta^\inn_6\,\A_{ab}^{12}\,\bB_{ab}^{134}\,\bB_{ab}^{156}-\notag\\
&-10\int\frac{\d\Omega_{1\dots6}}{(4\pi)^6}\,\Theta^\out_1\,\Theta^\inn_2\,\Theta^\inn_3\,\Theta^\out_4\,\Theta^\out_5\,\Theta^\inn_6\,\A_{ab}^{12}\,\bA_{ab}^{13}\,\bC_{ab}^{1456}+\notag\\
&+10\int\frac{\d\Omega_{1\dots6}}{(4\pi)^6}\,\Theta^\out_1\,\Theta^\out_2\,\Theta^\inn_3\,\Theta^\out_4\,\Theta^\out_5\,\Theta^\inn_6\,\B_{ab}^{123}\,\bC_{ab}^{1456}+\notag\\
&+10\int\frac{\d\Omega_{1\dots6}}{(4\pi)^6}\,\Theta^\out_1\,\Theta^\inn_2\,\Theta^\inn_3\,\Theta^\out_4\,\Theta^\inn_5\,\Theta^\inn_6\,\A_{ab}^{12}\,\bA_{ab}^{13}\,\bar{\mathfrak{A}}_{ab}^{1456}-\notag\\
&-10\int\frac{\d\Omega_{1\dots6}}{(4\pi)^6}\,\Theta^\out_1\,\Theta^\out_2\,\Theta^\inn_3\,\Theta^\out_4\,\Theta^\inn_5\,\Theta^\inn_6\,\B_{ab}^{123}\,\bar{\mathfrak{A}}_{ab}^{1456}+\notag\\
&+5 \int\frac{\d\Omega_{1\dots6}}{(4\pi)^6}\,\Theta^\out_1\,\Theta^\inn_2\,\Theta^\out_3\,\Theta^\inn_4\,\Theta^\inn_5\,\Theta^\inn_6\,\A_{ab}^{12}\,\bar{\mathbb{A}}_{ab}^{13456}-\notag\\
&-15\int\frac{\d\Omega_{1\dots6}}{(4\pi)^6}\,\Theta^\out_1\,\Theta^\inn_2\,\Theta^\out_3\,\Theta^\inn_4\,\Theta^\out_5\,\Theta^\inn_6\,\A_{ab}^{12}\,\bar{\tilde{\mathfrak{A}}}_{ab}^{13456}-\notag\\
&-5 \int\frac{\d\Omega_{1\dots6}}{(4\pi)^6}\,\Theta^\out_1\,\Theta^\inn_2\,\Theta^\out_3\,\Theta^\out_4\,\Theta^\inn_5\,\Theta^\inn_6\,\A_{ab}^{12}\,\bar{\mathfrak{B}}_{ab}^{13456}+\notag\\
&+5 \int\frac{\d\Omega_{1\dots6}}{(4\pi)^6}\,\Theta^\out_1\,\Theta^\inn_2\,\Theta^\out_3\,\Theta^\out_4\,\Theta^\out_5\,\Theta^\inn_6\,\A_{ab}^{12}\,\bar{\mathcal{D}}_{ab}^{13456}+\notag\\
&+  \int\frac{\d\Omega_{1\dots6}}{(4\pi)^6}\,\Theta^\out_1\,\Theta^\out_2\,\Theta^\inn_3\,\Theta^\inn_4\,\Theta^\inn_5\,\Theta^\inn_6\,\mathcal{J}_{ab}^{123456}-\notag\\
&-6 \int\frac{\d\Omega_{1\dots6}}{(4\pi)^6}\,\Theta^\out_1\,\Theta^\out_2\,\Theta^\out_3\,\Theta^\inn_4\,\Theta^\inn_5\,\Theta^\inn_6\,\mathcal{K}_{ab}^{123456}+\notag\\
&+3 \int\frac{\d\Omega_{1\dots6}}{(4\pi)^6}\,\Theta^\out_1\,\Theta^\out_2\,\Theta^\out_3\,\Theta^\inn_4\,\Theta^\out_5\,\Theta^\inn_6\,\mathcal{L}_{ab}^{123456}+\notag\\
&+4 \int\frac{\d\Omega_{1\dots6}}{(4\pi)^6}\,\Theta^\out_1\,\Theta^\out_2\,\Theta^\out_3\,\Theta^\out_4\,\Theta^\inn_5\,\Theta^\inn_6\,\mathcal{P}_{ab}^{123456}-\notag\\
&-4 \int\frac{\d\Omega_{1\dots6}}{(4\pi)^6}\,\Theta^\out_1\,\Theta^\out_2\,\Theta^\out_3\,\Theta^\inn_4\,\Theta^\inn_5\,\Theta^\inn_6\,\mathcal{Q}_{ab}^{123456}-\notag\\
&-  \int\frac{\d\Omega_{1\dots6}}{(4\pi)^6}\,\Theta^\out_1\,\Theta^\out_2\,\Theta^\out_3\,\Theta^\inn_4\,\Theta^\inn_5\,\Theta^\inn_6\,\mathbb{B}^{123456}_{ab}+\notag\\
&+3 \int\frac{\d\Omega_{1\dots6}}{(4\pi)^6}\,\Theta^\out_1\,\Theta^\out_2\,\Theta^\out_3\,\Theta^\inn_4\,\Theta^\out_5\,\Theta^\inn_6\,\tilde{\mathfrak{B}}^{123456}_{ab}+\notag\\
&+  \int\frac{\d\Omega_{1\dots6}}{(4\pi)^6}\,\Theta^\out_1\,\Theta^\out_2\,\Theta^\out_3\,\Theta^\out_4\,\Theta^\inn_5\,\Theta^\inn_6\,\mathfrak{C}^{123456}_{ab}\,,
\label{eq:S6}
\end{align}
where
\begin{subequations}
\begin{align}
\mathcal{J}_{ab}^{ijk\ell mn}& =w_{ab}^i\left(\A_{ai}^{jk}\bA_{ai}^{j\ell}\bA_{ai}^{jm}\bA_{ai}^{jn} +\A_{ib}^{jk}\bA_{ib}^{j\ell}\bA_{ib}^{jm}\bA_{ib}^{jn}-\A_{ab}^{jk}\bA_{ab}^{j\ell}\bA_{ab}^{jm}\bA_{ab}^{jn}\right),\\
\mathcal{K}_{ab}^{ijk\ell mn}&= w_{ab}^i\left(\B_{ai}^{jk\ell}\bA_{ai}^{jm}\bA_{ai}^{jn}+\B_{ib}^{jk\ell}\bA_{ib}^{jm}\bA_{ib}^{jn}-\B_{ab}^{jk\ell}\bA_{ab}^{jm}\bA_{ab}^{jn}\right),\\
\mathcal{L}_{ab}^{ijk\ell mn}&=w_{ab}^i\left(\B_{ai}^{jk\ell}\bB_{ai}^{jmn}+\B_{ib}^{jk\ell}\bB_{ib}^{jmn}-\B_{ab}^{jk\ell}\bB_{ab}^{jmn}\right),\\
\mathcal{P}_{ab}^{ijk\ell mn}&=w_{ab}^i\left(\A_{ai}^{jn}\bC_{ai}^{jk\ell m}+\A_{ib}^{jn}\bC_{ib}^{jk\ell m}-\A_{ab}^{jn}\bC_{ab}^{jk\ell m}\right),\\
\mathcal{Q}_{ab}^{ijk\ell mn}&=w_{ab}^i\left(\A_{ai}^{jn}\bar{\mathfrak{A}}_{ai}^{jk\ell m}+\A_{ib}^{jn}\bar{\mathfrak{A}}_{ib}^{jk\ell m}-\A_{ab}^{jn}\bar{\mathfrak{A}}_{ab}^{jk\ell m}\right).
\end{align}
\end{subequations}
At this point we emphasize that the unintegrated results $\S_{ab}^{(n)}$ that we have presented are in fact very general and applicable to the computation of the distribution of any non-global observable, at large $\Nc$. The evaluation of these integrals for the specific case of the hemisphere mass distribution will be presented in section 5. Before doing so we show, in the next section, how squared eikonal amplitudes for the emission of soft energy-ordered gluons may be extracted from the above expressions for $\S_{ab}^{(n)}$.

\section*{Eikonal amplitudes from the BMS equation}

The integrals involved in the expressions of each of the coefficients $\S_{ab}^{(n)}$ correctly reproduce the squared eikonal amplitudes for the emission of $n$ soft energy-ordered gluons at large $\Nc$. The phase space of these integrals encodes the impact of the ``measurement operator'', introduced in Ref. \cite{Schwartz:2014wha}, on the various squared amplitudes corresponding to all possible real/virtual gluon configurations at a given order $n$, including the possible angular configurations (inside or outside the measured region). Furthermore, the constant integers that appear in front of the integrals merely result from identical contributions of different permutations of the emitted gluons. Take, for instance, the four-loops term from Eq. \eqref{eq:S4}
\begin{equation}
3\int\frac{\d\Omega_{1234}}{(4\pi)^4}\,\Theta^\out_1\,\Theta^\inn_2\,\Theta^\out_3\,\Theta^\inn_4\,\A_{ab}^{12}\,\bB_{ab}^{134}\,.
\end{equation}
This term can be rewritten as
\begin{align}
\int\frac{\d\Omega_{1234}}{(4\pi)^4}&\left[\Theta^\out_1\,\Theta^\inn_2\,\Theta^\out_3\,\Theta^\inn_4\,\A_{ab}^{12}\,\bB_{ab}^{134}+
\Theta^\out_1\,\Theta^\out_2\,\Theta^\inn_3\,\Theta^\inn_4\,\A_{ab}^{13}\,\bB_{ab}^{124}+\right.\notag\\
&\left.+ \Theta^\out_1\,\Theta^\out_2\,\Theta^\inn_3\,\Theta^\inn_4\,\A_{ab}^{14}\,\bB_{ab}^{123}\right].
\end{align}
These three terms actually give identical results after integration, but each of them originates from a different term in the amplitude squared that comes from one of the permutations of the emitted gluons. The color factor associated with the squared amplitude that we extract at $n^{\mathrm{th}}$ order (at large $\Nc$) is $\mathrm{N_c^{\it n}}$. As stated before, we invoke the replacement $\mathrm{N_c^{\it n}}\to2\,\CF\,\mathrm{C_A^{{\it n}-1}}$ in order to partially restore the finite-$\Nc$ color structure of the squared amplitudes.

It should be emphasized, though, that the results presented in the previous section explicitly reproduce just what was referred to in Ref. \cite{Delenda:2015tbo} as the ``\emph{irreducible}'' parts of the squared amplitudes at a given order. The ``\emph{reducible}'' parts of the squared eikonal amplitudes are related to the interference terms that one obtains by expanding the proposed exponential solution \eqref{eq:proposed}. The reducible parts of the squared amplitudes at order $n$ are written purely in terms of squared amplitudes at previous orders $m<n$. Extracting these reducible amplitudes is trivial, and we shall show the results below.

In the remaining part of this section we present the emission squared amplitudes deduced from the exponential solution to the BMS equation up to six loops. We follow the notation of Ref. \cite{Delenda:2015tbo}, where $\W_{12\dots m}^{\scriptscriptstyle X}$ represents the $m$-gluon-emission amplitude squared and $X$ denotes the real-virtual configuration of the emitted gluons. For instance, $\W_{123}^{\R\V\R}$ is the squared amplitude of emission of three gluons with $k_1$ and $k_3$ being real and $k_2$ being virtual. We shall only present the real-emission amplitudes. The virtual corrections may readily be deduced from the latter as explained in Ref. \cite{Delenda:2015tbo}.

The factorized squared amplitude for the emission of a single soft gluon ($n=1$) off a dipole $(ab)$ is read from the expression of $\S_{ab}^{(1)}$ \eqref{eq:Sab1} to be
\begin{equation}
\W_1^\R = 2\,\CF\, w_{ab}^1\,.
\end{equation}
Notice that strictly speaking we have omitted a factor $g_s^2/\omega_1^2$, with $g_s^2=4\pi\as$ the strong coupling, which has been absorbed into the evolution parameter $t$. At two loops, the emission squared amplitude is given by
\begin{equation}
\W_{12}^{\R\R}=\W_1^\R\,\W_2^\R+\WI_{12}^{\R\R}\,,
\end{equation}
where the irreducible part is read from $\S_{ab}^{(2)}$ (last line of Eq. \eqref{eq:S2}) to be
\begin{equation}
\WI_{12}^{\R\R}=2\,\CF\,\CA\,\A_{ab}^{12}\,.
\end{equation}
At three loops we have
\begin{equation}
\W_{123}^{\R\R\R}=\W_1^\R\,\W_2^\R\,\W_3^\R+\W_1^\R\,\WI_{23}^{\R\R}+\W_2^\R\,\WI_{13}^{\R\R}+\W_3^\R\,\WI_{12}^{\R\R}+\WI_{123}^{\R\R\R}\,,
\end{equation}
with the totally irreducible component (read from Eq. \eqref{eq:S3})
\begin{equation}
\WI_{123}^{\R\R\R}=2\,\CF\,\mathrm{C_A^2}\left(\A_{ab}^{12}\,\bA_{ab}^{13}+\B_{ab}^{123}\right).
\end{equation}
We note here that the minus sign associated with the term $\B_{ab}^{123}$ in Eq. \eqref{eq:S3}, as well as the different phase space of integration in the two terms of this equation, results from consideration of virtual contributions. This is clearly explained in Ref. \cite{Khelifa-Kerfa:2015mma}. For instance, the contribution $\Theta^\out_1\,\Theta^\inn_2\,\Theta^\inn_3\,\A_{ab}^{12}\,\bA_{ab}^{13}$ is associated with the irreducible part of the squared amplitude of emission with gluon $k_2$ being virtual $\WI_{123}^{\R\V\R}$, where
\begin{equation}
\WI_{123}^{\R\V\R}=-2\,\CF\,\mathrm{C_A^2}\,\A_{ab}^{12}\,\bA_{ab}^{13}\,,
\end{equation}
while the contribution $\Theta^\out_1\,\Theta^\out_2\,\Theta^\inn_3\,\B_{ab}^{123}$ is associated with the sum of the two amplitudes $\WI_{123}^{\R\V\R}+\WI_{123}^{\R\R\R}$, as explained in Ref. \cite{Khelifa-Kerfa:2015mma}.

The four-loops amplitude squared is given by
\begin{align}
\W_{1234}^{\R\R\R\R}=&\,\W_1^\R\W_2^\R\W_3^\R\W_4^\R+\left(\W_1^\R\W_2^\R\WI_{34}^{\R\R}+\text{perm.}\right)+\left(\WI_{12}^{\R\R}\WI_{34}^{\R\R}+\text{perm.}\right)\notag\\
&+\left(\W_1^\R\WI_{234}^{\R\R\R}+\text{perm.}\right)+\WI_{1234}^{\R\R\R\R}+\text{finite-$\Nc$ contributions}\,,
\end{align}
where ``$\text{perm.}$'' stands for all possible permutations of the gluons that do not reproduce the same term twice (in order to avoid double counting). \footnote{When permuting the irreducible parts, e.g. $\WI_{ijk}^{\R\R\R}$, the indices must always be ordered such that $i<j<k$.} The irreducible part of the squared amplitude at this order is read from Eq. \eqref{eq:S4} to be
\begin{align}
\WI_{1234}^{\R\R\R\R}=2\,\CF\,\mathrm{C_A^3}&\left(\A_{ab}^{12}\,\bA_{ab}^{13}\,\bA_{ab}^{14}+\mathfrak{A}_{ab}^{1234}+\C_{ab}^{1234}+
\A_{ab}^{12}\,\bB_{ab}^{134}+\A_{ab}^{13}\,\bB_{ab}^{124}+\right.\notag\\
&\left.+\A_{ab}^{14}\,\bB_{ab}^{123}\right).
\end{align}

The five-loops squared amplitude is given by
\begin{align}
\W_{12345}^{\R\R\R\R\R}=&\,\W_1^\R\W_2^\R\W_3^\R\W_4^\R\W_5^\R+\left(\W_1^\R\W_2^\R\W_3^\R\WI_{45}^{\R\R}+\text{perm.}\right)+\notag\\
&+\left(\W_1^\R\WI_{23}^{\R\R}\WI_{45}^{\R\R}+\text{perm.}\right)+\left(\W_1^\R\W_2^\R\WI_{345}^{\R\R\R}+\text{perm.}\right)+\notag\\
&+\left(\WI_{12}^{\R\R}\WI_{345}^{\R\R\R}+\text{perm.}\right)+\left(\W_1^\R\WI_{2345}^{\R\R\R\R}+\text{perm.}\right)+\notag\\
&+\WI_{12345}^{\R\R\R\R\R}+\text{finite-$\Nc$ contributions}\,.
\end{align}
The irreducible contribution can be read from Eq. \eqref{eq:S5}
\begin{align}
\WI_{12345}^{\R\R\R\R\R}=2\,\CF\,\mathrm{C_A^4}&\Big(\D_{ab}^{12345}+\A_{ab}^{12}\,\bA_{ab}^{13}\,\bA_{ab}^{14}\,\bA_{ab}^{15}+\notag\\
&+\B_{ab}^{123}\,\bA_{ab}^{14}\,\bA_{ab}^{15}+\B_{ab}^{124}\,\bA_{ab}^{13}\,\bA_{ab}^{15}+\B_{ab}^{125}\,\bA_{ab}^{13}\,\bA_{ab}^{14}+\notag\\
&+\B_{ab}^{134}\,\bA_{ab}^{12}\,\bA_{ab}^{15}+\B_{ab}^{135}\,\bA_{ab}^{12}\,\bA_{ab}^{14}+\B_{ab}^{145}\,\bA_{ab}^{12}\,\bA_{ab}^{13}+\notag\\
&+\B_{ab}^{123}\,\bB_{ab}^{145}+\B_{ab}^{124}\,\bB_{ab}^{135}+\B_{ab}^{125}\,\bB_{ab}^{134}+\notag\\
&+\A_{ab}^{15}\,\bar{\mathfrak{A}}_{ab}^{1234}+\A_{ab}^{14}\,\bar{\mathfrak{A}}_{ab}^{1235}+\A_{ab}^{13}\,\bar{\mathfrak{A}}_{ab}^{1245}+\A_{ab}^{12}\,\bar{\mathfrak{A}}_{ab}^{1345}+\notag\\
&+\A_{ab}^{15}\,\bC_{ab}^{1234}+\A_{ab}^{14}\,\bC_{ab}^{1235}+\A_{ab}^{13}\,\bC_{ab}^{1245}+\A_{ab}^{12}\,\bC_{ab}^{1345}\notag\\
&+\tilde{\mathfrak{A}}_{ab}^{12345}+\tilde{\mathfrak{A}}_{ab}^{12435}+\tilde{\mathfrak{A}}_{ab}^{12534}+\mathbb{A}_{ab}^{12345}+\mathfrak{B}^{12345}_{ab}\Big).
\end{align}

At six loops we have
\begin{align}\label{eq:W6}
\W_{123456}^{\R\R\R\R\R\R}=&\,\W_1^\R \W_2^\R \W_3^\R \W_4^\R\W_5^\R\W_6^\R+\left(\W_1^\R \W_2^\R \W_3^\R\W_4^\R \WI_{56}^{\R\R}+\text{perm.}\right)+\notag\\
&+\left(\W_1^\R \W_2^\R  \WI_{34}^{\R\R} \WI_{56}^{\R\R}+\text{perm.}\right)+\left(\WI_{12}^{\R\R} \WI_{34}^{\R\R} \WI_{56}^{\R\R}+\text{perm.}\right)+\notag\\
&+\left(\W_1^\R\W_2^\R\W_3^\R\WI_{456}^{\R\R\R}+\text{perm.}\right)+\left(\W_1^\R \WI_{23}^{\R\R} \WI_{456}^{\R\R\R}+\text{perm.}\right)+\notag\\
&+\left( \WI_{123}^{\R\R\R}\WI_{456}^{\R\R\R}+\text{perm.}\right)+\left(\WI_{12}^{\R\R}\WI_{3456}^{\R\R\R\R}+\text{perm.}\right)+\notag\\
&+\left(\W_1^\R\W_2^\R\WI_{3456}^{\R\R\R\R}+\text{perm.}\right)+\left(\W_1^\R\WI_{23456}^{\R\R\R\R\R}+\text{perm.}\right)+\notag\\
&+\WI_{123456}^{\R\R\R\R\R\R}+\text{finite-$\Nc$ contributions}\,,
\end{align}
where the irreducible contribution at this order deduced from Eq. \eqref{eq:S6} is
\begin{align}
&\WI_{123456}^{\R\R\R\R\R}=\A_{ab}^{12}\bA_{ab}^{13}\bA_{ab}^{14}\bA_{ab}^{15}\bA_{ab}^{16}+\A_{ab}^{12}\bA_{ab}^{13}\bA_{ab}^{14}\bB_{ab}^{156}+\A_{ab}^{12}\bA_{ab}^{13}\bA_{ab}^{15}\bB_{ab}^{146}+\notag\\
&+\A_{ab}^{12}\bA_{ab}^{13}\bA_{ab}^{16}\bB_{ab}^{145}+\A_{ab}^{12}\bA_{ab}^{14}\bA_{ab}^{15}\bB_{ab}^{136}+\A_{ab}^{12}\bA_{ab}^{14}\bA_{ab}^{16}\bB_{ab}^{135}+\A_{ab}^{12}\bA_{ab}^{15}\bA_{ab}^{16}\bB_{ab}^{134}+\notag\\
&+\A_{ab}^{13}\bA_{ab}^{14}\bA_{ab}^{15}\bB_{ab}^{126}+\A_{ab}^{13}\bA_{ab}^{14}\bA_{ab}^{16}\bB_{ab}^{125}+\A_{ab}^{13}\bA_{ab}^{15}\bA_{ab}^{16}\bB_{ab}^{124}+\A_{ab}^{14}\bA_{ab}^{15}\bA_{ab}^{16}\bB_{ab}^{123}+\notag\\
&+\A_{ab}^{12}\bB_{ab}^{134}\bB_{ab}^{156}+\A_{ab}^{12}\bB_{ab}^{135}\bB_{ab}^{146}+\A_{ab}^{12}\bB_{ab}^{136}\bB_{ab}^{145}+\A_{ab}^{13}\bB_{ab}^{124}\bB_{ab}^{156}+\A_{ab}^{13}\bB_{ab}^{125}\bB_{ab}^{146}\notag\\
&+\A_{ab}^{13}\bB_{ab}^{126}\bB_{ab}^{145}+\A_{ab}^{14}\bB_{ab}^{123}\bB_{ab}^{156}+\A_{ab}^{14}\bB_{ab}^{125}\bB_{ab}^{136}+\A_{ab}^{14}\bB_{ab}^{126}\bB_{ab}^{135}+\A_{ab}^{15}\bB_{ab}^{123}\bB_{ab}^{146}\notag\\
&+\A_{ab}^{15}\bB_{ab}^{124}\bB_{ab}^{136}+\A_{ab}^{15}\bB_{ab}^{126}\bB_{ab}^{134}+\A_{ab}^{16}\bB_{ab}^{123}\bB_{ab}^{145}+\A_{ab}^{16}\bB_{ab}^{124}\bB_{ab}^{135}+\A_{ab}^{16}\bB_{ab}^{125}\bB_{ab}^{134}\notag\\
&+\A_{ab}^{12}\bA_{ab}^{13} \bC_{ab}^{1456}+\A_{ab}^{12}\bA_{ab}^{14} \bC_{ab}^{1356}+\A_{ab}^{12}\bA_{ab}^{15} \bC_{ab}^{1346}+\A_{ab}^{12}\bA_{ab}^{16} \bC_{ab}^{1345}+\A_{ab}^{13}\bA_{ab}^{14} \bC_{ab}^{1256}\notag\\
&+\A_{ab}^{13}\bA_{ab}^{15} \bC_{ab}^{1246}+\A_{ab}^{13}\bA_{ab}^{16} \bC_{ab}^{1245}+\A_{ab}^{14}\bA_{ab}^{15} \bC_{ab}^{1236}+\A_{ab}^{14}\bA_{ab}^{16} \bC_{ab}^{1235}+\A_{ab}^{15}\bA_{ab}^{16} \bC_{ab}^{1234}\notag\\
&+\B_{ab}^{123}\bC_{ab}^{1456}+\B_{ab}^{124}\bC_{ab}^{1356}+\B_{ab}^{125}\bC_{ab}^{1346}+\B_{ab}^{126}\bC_{ab}^{1345}+\B_{ab}^{134}\bC_{ab}^{1256}+\B_{ab}^{135}\bC_{ab}^{1246}+\notag\\
&+\B_{ab}^{136}\bC_{ab}^{1245}+\B_{ab}^{145}\bC_{ab}^{1236}+\B_{ab}^{146}\bC_{ab}^{1235}+\B_{ab}^{156}\bC_{ab}^{1234}+\A_{ab}^{12}\bA_{ab}^{13}\bar{\mathfrak{A}}_{ab}^{1456}+\notag\\
&+\A_{ab}^{12}\bA_{ab}^{14}\bar{\mathfrak{A}}_{ab}^{1356}+\A_{ab}^{12}\bA_{ab}^{15}\bar{\mathfrak{A}}_{ab}^{1346}+\A_{ab}^{12}\bA_{ab}^{16}\bar{\mathfrak{A}}_{ab}^{1345}+\A_{ab}^{13}\bA_{ab}^{14}\bar{\mathfrak{A}}_{ab}^{1256}+\notag\\
&+\A_{ab}^{13}\bA_{ab}^{15}\bar{\mathfrak{A}}_{ab}^{1246}+\A_{ab}^{13}\bA_{ab}^{16}\bar{\mathfrak{A}}_{ab}^{1245}+\A_{ab}^{14}\bA_{ab}^{15}\bar{\mathfrak{A}}_{ab}^{1236}+\A_{ab}^{14}\bA_{ab}^{16}\bar{\mathfrak{A}}_{ab}^{1235}+\notag\\
&+\A_{ab}^{15}\bA_{ab}^{16}\bar{\mathfrak{A}}_{ab}^{1234}+\B_{ab}^{123}\bar{\mathfrak{A}}_{ab}^{1456}+\B_{ab}^{124}\bar{\mathfrak{A}}_{ab}^{1356}+\B_{ab}^{125}\bar{\mathfrak{A}}_{ab}^{1346}+\B_{ab}^{126}\bar{\mathfrak{A}}_{ab}^{1345}+\notag\\
&+\B_{ab}^{134}\bar{\mathfrak{A}}_{ab}^{1256}+\B_{ab}^{135}\bar{\mathfrak{A}}_{ab}^{1246}+\B_{ab}^{136}\bar{\mathfrak{A}}_{ab}^{1245}+\B_{ab}^{145}\bar{\mathfrak{A}}_{ab}^{1236}+\B_{ab}^{146}\bar{\mathfrak{A}}_{ab}^{1235}+\B_{ab}^{156}\bar{\mathfrak{A}}_{ab}^{1234}\notag\\
&+\A_{ab}^{12}\bar{\mathbb{A}}_{ab}^{13456}+\A_{ab}^{13}\bar{\mathbb{A}}_{ab}^{12456}+\A_{ab}^{14}\bar{\mathbb{A}}_{ab}^{12356}+\A_{ab}^{15}\bar{\mathbb{A}}_{ab}^{12346}+\A_{ab}^{16}\bar{\mathbb{A}}_{ab}^{12345}+\notag\\
&+\A_{ab}^{12}\bar{\tilde{\mathfrak{A}}}_{ab}^{13456}+\A_{ab}^{12}\bar{\tilde{\mathfrak{A}}}_{ab}^{13546}+\A_{ab}^{12}\bar{\tilde{\mathfrak{A}}}_{ab}^{13645}+\A_{ab}^{13}\bar{\tilde{\mathfrak{A}}}_{ab}^{12456}
+\A_{ab}^{13}\bar{\tilde{\mathfrak{A}}}_{ab}^{12546}+\A_{ab}^{13}\bar{\tilde{\mathfrak{A}}}_{ab}^{12645}\notag\\
&+\A_{ab}^{14}\bar{\tilde{\mathfrak{A}}}_{ab}^{12356}+\A_{ab}^{14}\bar{\tilde{\mathfrak{A}}}_{ab}^{12536}+\A_{ab}^{14}\bar{\tilde{\mathfrak{A}}}_{ab}^{12635}
+\A_{ab}^{15}\bar{\tilde{\mathfrak{A}}}_{ab}^{12346}+\A_{ab}^{15}\bar{\tilde{\mathfrak{A}}}_{ab}^{12436}+\A_{ab}^{15}\bar{\tilde{\mathfrak{A}}}_{ab}^{12634}\phantom{+}\notag\\
&+\A_{ab}^{16}\bar{\tilde{\mathfrak{A}}}_{ab}^{12345}+\A_{ab}^{16}\bar{\tilde{\mathfrak{A}}}_{ab}^{12435}
+\A_{ab}^{16}\bar{\tilde{\mathfrak{A}}}_{ab}^{12534}+\notag\\
&+\A_{ab}^{12} \bar{\mathfrak{B}}_{ab}^{13456}+\A_{ab}^{13} \bar{\mathfrak{B}}_{ab}^{12456}+\A_{ab}^{14} \bar{\mathfrak{B}}_{ab}^{12356}+\A_{ab}^{15} \bar{\mathfrak{B}}_{ab}^{12346}+\A_{ab}^{16} \bar{\mathfrak{B}}_{ab}^{12345}+\notag\\
&+\A_{ab}^{12} \bar{\mathcal{D}}_{ab}^{13456}+\A_{ab}^{13} \bar{\mathcal{D}}_{ab}^{12456}+\A_{ab}^{14} \bar{\mathcal{D}}_{ab}^{12356}+\A_{ab}^{15} \bar{\mathcal{D}}_{ab}^{12346}+\A_{ab}^{16}\bar{\mathcal{D}}_{ab}^{12345}+\notag\\
&+\mathcal{K}_{ab}^{123456}+\mathcal{K}_{ab}^{123546}+\mathcal{K}_{ab}^{123645}+\mathcal{K}_{ab}^{124536}+\mathcal{K}_{ab}^{124635}+\mathcal{K}_{ab}^{125634}+\notag\\
&+\mathcal{L}_{ab}^{123456}+\mathcal{L}_{ab}^{123546}+\mathcal{L}_{ab}^{123645}+\mathcal{P}_{ab}^{123456}+\mathcal{P}_{ab}^{123465}+\mathcal{P}_{ab}^{123564}+\mathcal{P}_{ab}^{124563}+\notag\\
&+\mathcal{Q}_{ab}^{123456}+\mathcal{Q}_{ab}^{123465}+\mathcal{Q}_{ab}^{123564}+\mathcal{Q}_{ab}^{124563}+\tilde{\mathfrak{B}}^{123456}_{ab}+\tilde{\mathfrak{B}}^{123546}_{ab}+\tilde{\mathfrak{B}}^{123645}_{ab}+\notag\\
&+\mathcal{J}_{ab}^{123456}+\mathbb{B}^{123456}_{ab}+\mathfrak{C}^{123456}_{ab}+\E_{ab}^{123456}\,.\label{eq:Ird6}
\end{align}

Notice how the integer coefficient that multiplies each integral in Eq. \eqref{eq:S6} exactly corresponds to the number of possible permutations of gluons in the squared amplitude. The result at six loops has not been previously reported in the literature, and we have deduced it here from the BMS equation. Furthermore, we have verified this result by comparing it to the output of the \texttt{Mathematica} program \texttt{EikAmp} \cite{Delenda:2015tbo}. Recall that \texttt{EikAmp} additionally produces terms that are subleading in color, i.e., finite-$\Nc$ contributions, and thus provides squared amplitudes that are more accurate in terms of color structure.

\section*{NGLs in the hemisphere mass distribution}

We present in this section the results of integrations for the coefficients $\S_{ab}^{(n)}$ of NGLs up to $n=4$, in the case of back-to-back di-jet production events in $e^+e^-$ collisions, where we measure the invariant mass of the hemisphere defined by one of the jets. The iterative structure of the integrals suggests the use of Goncharov polylogarithms (GPLs), symbols and co-product machinery \cite{Goncharov2001MultiplePA, Goncharov1998MultiplePC}, which greatly simplifies the analytical computations of the said integrations \cite{Schwartz:2014wha}. The results we report herein serve as confirmation of the semi-analytical calculations carried out in Ref. \cite{Khelifa-Kerfa:2015mma}. We leave the details of integrations to \ref{app:NGLs-Calculaions} and confine ourselves here to stating the full result up to fourth order.

The resummed hemisphere mass distribution may be expressed as follows
\begin{equation}
\sigma(\rho)=\sigma^{\mathrm{p}}(\rho)\,\times\,\sigma^{\NG}(\rho)\,,
\end{equation}
where $\sigma^{\mathrm{p}}(\rho)$ is the primary Sudakov form factor given by \cite{Catani:1992ua}
\begin{equation}
\sigma^{\mathrm{p}}(\rho)=\frac{1}{\Gamma\left[1+\mathcal{R}'(\rho)\right]}\,\exp\left[-\mathcal{R}(\rho)-\gamma_E\,\mathcal{R}'(\rho)\right],
\end{equation}
with $\gamma_E\approx0.577$ being the Euler-Mascheroni constant, and for completeness the expression of the global radiator $\mathcal{R}$ is presented in \ref{app:rad}. The term $\sigma^{\NG}(\rho)$ is the resummed non-global factor given by
\begin{align}\label{eq:sigNG-exp}
\sigma^\NG(\rho)=\exp&\left(-\frac{(\Nc\,t)^2}{2!}\,\frac{\zeta_{2}}{2}+\frac{(\Nc\,t)^3}{3!}\,\frac{\zeta_{3}}{2}-\frac{(\Nc\,t)^4}{4!}\,\frac{29\,\zeta_{4}}{16}+\right.\notag\\
&+\left.\frac{(\Nc\,t)^5}{5!}\left[\frac{17}{4}\,\zeta_5+\frac{1}{2}\,\zeta_2\,\zeta_3\right]+\mathcal{O}(t^6)\right),
\end{align}
where the five-loops coefficient has been deduced from previous results in the literature \cite{Schwartz:2014wha}. The fixed-order expansion of this exponential gives
\begin{align}
\sigma^\NG(\rho)=&\,1-\frac{\pi^2}{24}\,(\Nc\,t)^2+\frac{\zeta_{3}}{12}\,(\Nc\,t)^3+\frac{\pi^4}{34\,560}\,(\Nc\,t)^4+\notag\\
&+\left(-\frac{\pi^2}{360}\,\zeta_3+\frac{17}{480}\,\zeta_5\right)(\Nc\,t)^5+\mathcal{O}(t^6)\,,
\end{align}
which confirms the results obtained in Refs. \cite{Schwartz:2014wha,Khelifa-Kerfa:2015mma}.

Since we have not computed higher NGLs coefficients, we can make a crude estimate of how large they may be. We do so by fitting the exponential solution \eqref{eq:proposed} truncated at seventh order to the full numerical resummation of NGLs obtained from the Monte Carlo dipole-evolution code of Ref. \cite{Dasgupta:2001sh}. The fitting values that we have obtained are
\begin{subequations}\label{eq:S6-S7-fit}
\begin{align}
\S_{ab}^{(6)}=&-13.34\,,\\
\S_{ab}^{(7)}=&+15.03\,.
\end{align}
\end{subequations}
The numerical values for the coefficients $\S_{ab}^{(n)}/n!$ that multiply $(\Nc\,t)^n$ in Eq. \eqref{eq:sigNG-exp} up to fifth order ($n=5$) are shown in Table \ref{tab1}. Also shown in the same table are the estimated coefficients at six and seven loops based on the fitting values \eqref{eq:S6-S7-fit}. We note that for typical phenomenological studies we have $\Nc\,t\lesssim 1$. Combined with the observation that the numerical values for the coefficients shown in Table \ref{tab1} become smaller at each higher order, it is expected that the series in the exponent in Eq. \eqref{eq:sigNG-exp} should converge fairly quickly.
\begin{table}[ht]
\caption{\label{tab1} The coefficients $\S_{ab}^{(n)}/n!$ multiplying $(\Nc\,t)^n$ at order $n$.}
\begin{center}
\begin{tabular}{ll}
\hline\noalign{\smallskip}
$n\hspace{5pt}$ & $\S_{ab}^{(n)}/n!$ \\
\noalign{\smallskip}\hline\noalign{\smallskip}
$2$ & $-0.41$ \\
$3$ & $+0.10$ \\
$4$ & $-0.082$ \\
$5$ & $+0.045$ \\
$6$ & $-0.018$ \\
$7$ & $+0.003$ \\
\noalign{\smallskip}\hline\noalign{\smallskip}
\end{tabular}
\end{center}
\end{table}

\subsection*{Two-loops ladder resummation}
\label{sec:2loopsres}

In this subsection we show how a class of terms that appear at each order in the perturbative expansion of NGLs may be resummed in the exponent of Eq. \eqref{eq:proposed} to all orders. Such a class of terms, dubbed ``ladder'' terms \cite{Delenda:2015tbo}, \footnote{The Feynman diagrams corresponding to these terms look like a ladder. See Fig. \ref{fig:ladder}.} seems to exhibit a symmetry pattern and starts at two loops by the expression \eqref{eq:S2}
\begin{equation}
\S_{ab}^{(2)}=-\int\frac{\d\Omega_{12}}{(4\pi)^2}\,\Theta^\out_1\,\Theta^\inn_2\,\A_{ab}^{12}\,.
\end{equation}
At higher orders ($n\geq 2$) the ladder terms appear as
\begin{align}
(-1)^{n-1}\int\frac{\d\Omega_{12\dots n}}{(4\pi)^n}\,\Theta^\out_1\,\omega_{ab}^1\prod_{i=2}^n\Theta^\inn_i\,\bA_{ab}^{1i}\,.
\end{align}
They may be represented by Feynman diagrams such as that shown in Fig. \ref{fig:ladder} at five loops.
\begin{figure}[ht]
\begin{center}
\includegraphics[width=.2\textwidth]{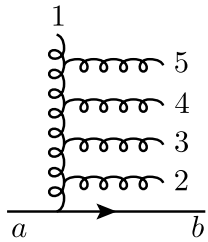}
\vspace{-3mm}
\caption{Diagrammatic representation of the ladder term at fifth order.}
\labelf{fig:ladder}
\end{center}
\vspace{-5mm}
\end{figure}
The result of integration of a given ladder term at order $n$ is given by the formula
\begin{equation}
(-1)^{n-1}\,(n-1)!\,\frac{\zeta_n}{2}\,,\qquad n\geq 2\,.
\end{equation}
Summing these terms to all orders in the exponent yields the result
\begin{align}\label{eq:Resum-Ladder}
\sigma_{\mathrm{ladder}}^\NG(\rho)&=\exp\left[\sum_{n=2}^\infty(-1)^{n-1}\,(n-1)!\,\frac{\zeta_n}{2}\,\frac{(\Nc\,t)^n}{n!}\right]\notag\\
&=\frac{1}{\sqrt{\Gamma\left(1+\Nc\,t\right)}}\,\exp\left[-\frac{\gamma_E}{2}\,\Nc\,t\right].
\end{align}
An identical result to the above was derived in Ref. \cite{Schwartz:2014wha} by means of solving the BMS equation as a differential equation up to two loops while ignoring higher-loop terms. Eq. \eqref{eq:Resum-Ladder} also accounts for the first radiator $R^{(1)}_{ab}(t)$ in Eq. (5.3) of Ref. \cite{Banfi:2002hw}. \footnote{Notice that the evolution parameter $t$ is denoted as $\delta$ in Ref. \cite{Banfi:2002hw}.} This means that the exponential solution \eqref{eq:proposed} may be factored out into a product of an infinite number of exponentials (or equivalently a sum of an infinite number of terms in the exponent), each of which resums a specific class of terms that exhibit a symmetry pattern. What we have computed above, i.e., ladder terms, is just the first obvious class of such terms. The resummation of less trivial classes of terms will be postponed to future work.

We show in Fig. \ref{fig:plot} plots of the ratio $\sigma^\NG/\sigma^\mathrm{DS}$, where $\sigma^\mathrm{DS}$ is a parameterization function that was obtained in Ref. \cite{Dasgupta:2001sh} by fitting to the output of a Monte Carlo dipole-evolution code developed therein to resum NGLs at large $\Nc$. It is given by
\begin{equation}
\sigma^{\mathrm{DS}}(\rho)=\exp\left[-\CF\,\CA\,\frac{\pi^2}{12}\,\frac{1+(0.85\,\CA\,t/2)^2}{1+(0.86\,\CA\,t/2)^{1.33}}\,t^2\right].
\end{equation}
\begin{figure}[ht]
\begin{center}
\includegraphics[width=.6\textwidth]{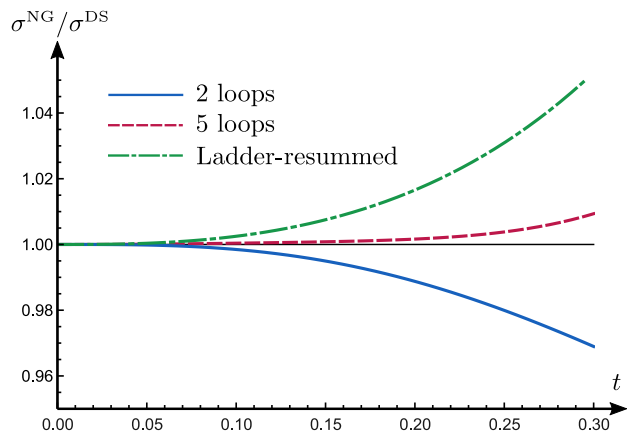}
\vspace{-3mm}
\caption{Plots of the ratio $\sigma^{\NG}/\sigma^{\mathrm{DS}}$ as a function of $t$ including terms in Eq. \eqref{eq:sigNG-exp} up to two loops, up to five loops, and including the ladder-resummed factor \eqref{eq:Resum-Ladder}.}
\labelf{fig:plot}
\end{center}
\vspace{-5mm}
\end{figure}
We show in Fig. \ref{fig:plot} three cases for $\sigma^\NG$. In one case we truncate the series in the exponent \eqref{eq:sigNG-exp} at two loops keeping only the leading term in the exponent. We notice that the exponential result $\sigma^\NG$ differs from the fully resummed result $\sigma^{\mathrm{DS}}$ only at the level of a maximum of $3\%$ for values of $t$ up to $0.3$ (which is equivalent to a value of $L=\ln(1/\rho)=5.3$ or $\rho=0.005$). In the second case we truncate the series in the exponent at fifth order. Here we see that the five-loops result performs better than the two-loops result with discrepancy less than $1\%$ for values of $t$ up to $0.3$. This indicates that adding a few more terms in the exponent one should be able to obtain a good agreement with the full numerical resummation of NGLs $\sigma^{\mathrm{DS}}$. In the third case we plot the ladder-resummed factor \eqref{eq:Resum-Ladder} which, on the other hand, does not seem to perform any better than the two-loops case. This indicates that it is necessary to resum higher-loop terms/classes in order to obtain a reliable analytical result that matches the fully-resummed numerical data.

\section*{Conclusions}

In the current work we have shown how to extract the squared eikonal amplitudes for the emission of soft energy-ordered gluons at large $\Nc$ from the non-linear integro-differential BMS equation. We have provided the explicit expressions for these squared amplitudes up to sixth order in the strong coupling. These amplitudes have actually been deduced from the formulae of the coefficients $\S^{(n)}_{ab}$ of NGLs, for which the BMS equation was initially developed. The expressions of $\S^{(n)}_{ab}$ that we provided are in fact general and may be applied to the resummation of any non-global QCD observable by simply performing the relevant integrations.

Moreover, we have explicitly verified that the squared amplitudes extracted from the BMS equation coincide with those presented in our previous work \cite{Delenda:2015tbo} in the large-$\Nc$ limit. The results of the latter reference were obtained by means of a \texttt{Mathematica} code that implements the dipole formalism in the eikonal approximation and captures the full color dependence of the amplitudes.

We additionally carried out analytical evaluations of the various integrals relevant to the NGLs coefficients $\S^{(n)}_{ab}$ up to fourth order for the specific case of the hemisphere mass distribution. We have thus confirmed the semi-numerical calculations presented previously in Ref. \cite{Khelifa-Kerfa:2015mma}. The fifth and sixth-order computations are quite delicate and will be presented in future publications. Furthermore, we compared our results to the full numerical resummation of Ref. \cite{Dasgupta:2001sh} and found that adding more and more terms in the exponent of the solution \eqref{eq:sigNG-exp} leads to better agreement with the Monte Carlo resummed result. Nonetheless, and as previously pointed out in the literature, the truncation of the series in the exponent at two loops already gives a good approximation for the full resummation for values of $t$ up to $0.3$.

We have also elaborated on the observation that the solution to the BMS equation may be represented by a product of an infinite number of exponential factors each of which resumming a class of terms contributing to the NGLs form factor. We have computed the first of such resummed factors, which corresponds to a class of terms whose Feynman diagrams resemble a ladder shape. It turns out, however, that this ladder resummation is not sufficient and higher-loop terms (e.g. ``cascade'' terms) cannot be neglected and should be computed for a precise and reliable solution to the BMS equation.

The current work may be extended in various ways. These include, to name a few, (a) analytically computing the fifth and sixth-order NGLs coefficients, (b) considering other non-global jet-shape distributions up to sixth order, (c) exploring the effect of jet algorithms on NGLs up to sixth order, and (d) exploiting new developments in mathematics especially in the field of non-linear integro-differential equations in the hope of finding a full analytical solution to the BMS equation. The latter may be possible given that the BMS equation is analogous to some well-known equations in physics, such as the Batlisky-Kovchegov (BK) equation (see Ref. \cite{Hatta:2013iba} and references therein), that have been thoroughly studied for a long time. We hope to address some of theses issues in the near future.

\section*{Acknowledgements}

This work is supported by:
\begin{itemize}
\item Deanship of Research at the Islamic University of Madinah (research project No. 40/107)
\item PRFU: B00L02UN050120190001 (Algeria)
\end{itemize}

Conflict of Interest: The authors declare that they have no conflicts of interest.

\appendix

\section*{Calculation of NGLs coefficients}
\label{app:NGLs-Calculaions}

\subsection*{One and two-loops calculations}

We evaluate in this section the integral \eqref{eq:Sab1} which gives the one-loop Sudakov coefficient $\S_{ij}^{(1)}$ for arbitrary dipole $(ij)$. We introduce the following brackets
\begin{subequations}
\begin{align}
\{ij\}&=1+c_i\,c_j+s_i\,s_j\cos\phi_{ij}\,,\\
[ij]  &=1+c_i\,c_j-s_i\,s_j\cos\phi_{ij}\,,
\end{align}
\end{subequations}
where we remind the reader that $c_i=\cos\theta_i$, $s_i=\sin\theta_i$, and $\phi_{ij}=\phi_i-\phi_j$. First, we carry out the azimuthal average
\begin{equation}
\I_{ij}(c_k)=\int_0^{2\pi}\frac{\d\phi_k}{2\pi}\,\frac{(ij)}{(ik)\,(jk)}\,,
\end{equation}
using contour integration techniques to obtain
\begin{equation}
\I_{ij}(c_k)=\I^{(1)}_{ij}(c_k)\left[\Theta(c_k-c_i)-\Theta(c_j-c_k)\right]+\I^{(2)}_{ij}(c_k)\left[\Theta(c_k-c_i)-\Theta(c_k-c_j)\right],
\end{equation}
with
\begin{subequations}
\begin{align}
\I^{(1)}_{ij}&=\frac{(ij)}{\left[ij\right]-2\left(c_i+c_j\right)c_k+\{ij\}c_k^2}\left[\frac{1-c_ic_k}{c_k-c_i}+\frac{1-c_jc_k}{c_k-c_j}\right],\\
\I^{(2)}_{ij}&=\frac{1}{1-c_k^2}\left[\frac{1-c_ic_k}{c_k-c_i}-\frac{1-c_jc_k}{c_k-c_j}\right].
\end{align}
\end{subequations}
Evaluating the polar integration results in a collinear divergence, as explained in the main text (see Ref. \cite{Khelifa-Kerfa:2015mma} for details).

The two-loops integral which represents the leading NGLs coefficient reads
\begin{equation}
\S_{ij}^{(2)}=-\int_{-1}^0\frac{\d c_{\ell}}{2}\,\frac{\d\phi_\ell}{2\pi}\,\frac{(ij)}{(i\ell)(j\ell)}\int_{0}^{1}\frac{\d c_k}{2}\left[\I_{i\ell}(c_k)+\I_{\ell j}(c_k)-\I_{ij}(c_k)\right].
\end{equation}
For the case $(ij)=(aj)$, with $c_j<0$ and $c_a=1$, one simply finds
\begin{equation}
\S_{aj}^{(2)}=-\frac{\zeta_2}{2}\,.
\end{equation}
Moreover, for the simpler case $(ij)=(ab)$, with $c_a=1$ and $c_b=-1$, one clearly has
\begin{equation}
\S_{ab}^{(2)}=-\frac{\zeta_2}{2}\,.
\end{equation}
Finally, for $(ij)=(ib)$, with $i$ an arbitrary leg outside the measured hemisphere ($c_i<0$), one has
\begin{equation}\label{5}
\S_{ib}^{(2)}=-\frac{1}{2}\left(\zeta_2-\Li_2\left[\frac{2c_i}{c_i-1}\right]\right).
\end{equation}

\subsection*{Three-loops calculations}

The three-loops NGLs coefficient for the hemisphere mass distribution in di-jet events in $e^+e^-$ collisions is given by
\begin{align}\label{6}
\S^{(3)}_{ab}=&\int_{-1}^0\frac{\d c_1}{2}\int_0^{2\pi}\frac{\d\phi_1}{2\pi}\int_0^1\frac{\d c_2}{2}\int_0^{2\pi}\frac{\d\phi_2}{2\pi}\int_0^1\frac{\d c_3}{2}\int_0^{2\pi}\frac{\d\phi_3}{2\pi}\,\A_{ab}^{12}\,\bar{\A_{ab}^{13}} - \notag\\
&-\int_{-1}^0\frac{\d c_1}{2}\int_0^{2\pi}\frac{\d\phi_1}{2\pi}\int_{-1}^0\frac{\d c_2}{2}\int_0^{2\pi}\frac{\d\phi_2}{2\pi}\int_0^1\frac{\d c_3}{2}\int_0^{2\pi}\frac{\d\phi_3}{2\pi}\,\B_{ab}^{123}\,.
\end{align}
For the ladder (first) term, which was also evaluated and resummed to all orders in the main text, we have the result
\begin{align}
I^{(3)}_1=\zeta_3\,.
\end{align}
For the cascade (second) term we have
\begin{align}
I^{(3)}_2=&-\int_{-1}^0\frac{\d c_1}{2}\int_0^{2\pi}\frac{\d\phi_1}{2\pi}\int_{-1}^0\frac{\d c_2}{2}\int_0^{2\pi}\frac{\d\phi_2}{2\pi}\times\notag\\
&\times\int_0^1\frac{\d c_3}{2}\int_0^{2\pi}\frac{\d\phi_3}{2\pi}\,w^1_{ab}\left(\A^{23}_{a1}+\A^{23}_{1b}-\A^{23}_{ab}\right).
\end{align}
Substituting the results of integration from the previous subsection we find
\begin{align}
I^{(3)}_2&=\int_{-1}^0\frac{\d c_1}{2}\int_0^{2\pi}\frac{\d\phi_1}{2\pi}\,w^1_{ab}\left(\S^{(2)}_{a1}+\S^{(2)}_{b1}-\S^{(2)}_{ab}\right)\notag\\
&=-\int_{-1}^0\frac{\d c_1}{2}\,\frac{2}{1-c_1^2}\left(\frac{\zeta_2}{2}-\frac{1}{2}\Li_2\left[\frac{2c_1}{c_1-1}\right]\right).
\end{align}
Making the change of variables
\begin{equation}
x=\frac{1+c_1}{1-c_1}\Rightarrow c_1=\frac{x-1}{x+1}\,,
\end{equation}
the cascade integral then reads
\begin{align}\label{7}
I^{(3)}_2&=-\frac{1}{4}\int_0^1\left[\zeta_2-\Li_2(1-x)\right]\d\ln x\,.
\end{align}
Letting $u=\ln x$ and using the dilogarithm identity
\begin{equation}
\Li_2(1-x)=\zeta_2-\ln x\ln(1-x)-\Li_2(x)\,,
\end{equation}
we obtain
\begin{align}
I_2^{(3)}&=-\frac{1}{4}\int_{-\infty}^0\left(u\ln(1-e^u)+\Li_2(e^u)\right)\d u\notag\\
&=-\frac{\zeta_3}{2}\,.
\end{align}
We thus have
\begin{equation}
\S^{(3)}_{ab}=\frac{\zeta_3}{2}\,.
\end{equation}
Note that we have checked this result numerically.

\subsection*{Four-loops calculations}

The NGLs coefficient at this order reads
\begin{align}\label{10}
\S_{ab}^{(4)}=&\int\frac{\d\Omega_{1234}}{(2\pi)^4}\,\Theta^\out_1\,\Theta^\inn_4\left(-\Theta^\inn_2\,\Theta^\inn_3\,\A^{12}_{ab}\,\bA^{13}_{ab}\,\bA^{14}_{ab}+\right.\notag\\
&\left.+3\,\,\Theta^\inn_2\,\Theta^\out_3\,\A^{12}_{ab}\,\bB^{134}_{ab}+\Theta^{\out}_2\,\Theta^\inn_3\,\mathfrak{A}^{1234}_{ab}-\Theta^\out_2\,\Theta^\out_3\,\C^{1234}_{ab}\right).
\end{align}
The ladder term is simple and yields the result
\begin{align}
I_1^{(4)}&=-\int\frac{\d\Omega_{1234}}{(2\pi)^4}\,\Theta^\out_1\,\Theta^\inn_2\,\Theta^\inn_3\,\Theta^\inn_4\,\A^{12}_{ab}\,\bA^{13}_{ab}\,\bA^{14}_{ab}\notag\\
&=-\int_{-1}^0\frac{\d c_1}{1-c_1^2}\,\ln^3\left(\frac{c_1-1}{2c_1}\right)\notag\\
&=-3\,\zeta_4\,.
\end{align}
The ladder-cascade term is given by
\begin{align}\label{11}
I^{(4)}_2&=3\int\frac{\d\Omega_{1234}}{(2\pi)^4}\,\Theta^\out_1\,\Theta^\inn_2\,\Theta^\out_3\,\Theta^\inn_4\,\A^{12}_{ab}\,\bB^{134}_{ab}\notag\\
&=\frac{3}{2}\int_{-1}^0\frac{\d c_1}{1-c_1^2}\,\ln\left(\frac{c_1-1}{2c_1}\right)\left(\zeta_2-\Li_2\left[\frac{2c_1}{c_1-1}\right]\right)\notag\\
&=\frac{21}{16}\,\zeta_4\,.
\end{align}
Note that we have used the Hopf algebra of co-products to carry out the above integral. Additionally we have
\begin{align}\label{12}
I_3^{(4)}&=\int\frac{\d\Omega_{1234}}{(2\pi)^4}\,\Theta^\out_1\,\Theta^\out_2\,\Theta^\inn_3\,\Theta^\inn_4\,\mathfrak{A}^{1234}_{ab}\notag\\
&=\frac{1}{4}\int\frac{\d\Omega_{12}}{(2\pi)^2}\,\Theta^\out_1\,\Theta^\out_2\,\frac{2}{1-c_1^2}\left[\frac{2(1-c_1c_2)}{(1-c^2_2)(12)}
\left(\ln\left[\frac{c_2-1}{2c_2^2(c_1-1)}\right]+\ln[12]\right)^2-\right.\notag\\
&\qquad\left.-\frac{8}{1-c_2^2}\ln^2\left(\frac{c_2-1}{2c_2}\right)\right]\notag\\
&=\frac{17}{16}\,\zeta_4\,.
\end{align}

The last integral to perform at four loops is the cascade term
\begin{align}
I^{(4)}_4&=-\int\frac{\d\Omega_{1234}}{(2\pi)^4}\,\Theta^\out_1\,\Theta^\out_2\,\Theta^\out_3\,\Theta^\inn_4\,\C^{1234}_{ab}\notag\\
&=-\int\frac{\d\Omega_{1234}}{(2\pi)^4}\,\Theta^\out_1\,\Theta^\out_2\,\Theta^\out_3\,\Theta^\inn_4\,w^1_{ab}\left[\B^{234}_{a1}+\B^{234}_{1b}-\B^{234}_{ab}\right].
\end{align}
Note that the integrations over each separate term is divergent, but the overall result is finite. We can put a spurious collinear cutoff $\epsilon$ on the integration over $c_1$ and perform each integral separately. The divergences cancel in the sum and $\epsilon$ disappears. The integral over the term involving $\B_{ab}^{123}$ is straightforward and yields the result
\begin{align}
I^{(4)}_{4,1}&=\int_{-1}^0\frac{\d c_1}{2}\int_0^{2\pi}\frac{\d\phi_1}{2\pi}\,w^1_{ab}\frac{\zeta_3}{2}=-\frac{\zeta_3}{4}\lim_{\epsilon\to0}\ln\frac{\epsilon}{2}.
\end{align}
The second integral is also easy and gives
\begin{align}
&I^{(4)}_{4,2}=-\int_{-1}^0\frac{\d c_1}{2}\int_0^{2\pi}\frac{\d\phi_1}{2\pi}\int_{-1}^0\frac{\d c_2}{2}\int_0^{2\pi}\frac{\d\phi_2}{2\pi}\,w^1_{ab}\,w^2_{1b}\,\frac{1}{2}\times\notag\\
&\qquad\times\left(\Li_2\left[\frac{2c_1}{c_1-1}\right]-\Li_2\left[\frac{2c_2}{c_2-1}\right]\right)\notag\\
&=\,-\frac{1}{4}\int_{-1}^0\frac{\d c_1}{1-c_1^2}\int_{-1}^0\frac{\d c_2}{1-c_2^2}\,\frac{(1+c_1)(1-c_2)}{|c_1-c_2|}\left(\Li_2\left[\frac{2c_1}{c_1-1}\right]-\Li_2\left[\frac{2c_2}{c_2-1}\right]\right).
\end{align}
Thus we obtain
\begin{equation}
I^{(4)}_{4,2}=-\frac{\zeta_3}{4}\lim_{\epsilon\to 0}\ln\frac{\epsilon}{2}-\frac{3\,\zeta_4}{4}\,.
\end{equation}

The remaining integral is the least trivial of all. It reads
\begin{align}
I^{(4)}_{4,3}&=-\int_{-1}^0\frac{\d c_1}{2}\int_0^{2\pi}\frac{\d\phi_1}{2\pi}\int_{-1}^0\frac{\d c_2}{2}\int_0^{2\pi}\frac{\d\phi_2}{2\pi}\,w^{1}_{ab}\left(w^{2}_{a1}+w^{2}_{1b}\right)\A^{\overline{34}}_{12}\notag\\
&=-\int_{-1}^0\frac{\d c_1}{1-c_1^2}\int_{-1}^0\frac{\d c_2}{1-c_2^2}\,(1-c_1c_2)\int_0^{2\pi}\frac{\d\phi_2}{2\pi}\,\frac{1}{(12)}\,\A^{\overline{34}}_{12}\,,
\end{align}
where the bar in $\A^{\overline{34}}_{12}$ means that both particles $3$ and $4$ have been integrated out with $3$ being out and $4$ being inside the measured hemisphere. We find for this term the result
\begin{equation}
I^{(4)}_{4,3}=\frac{\zeta_3}{2}\,\lim_{\epsilon\to0}\ln\frac{\epsilon}{2}-\frac{7\,\zeta_4}{16}\,.
\end{equation}
Thus the overall cascade contribution to the NGLs coefficient at four loops is given by
\begin{equation}
I^{(4)}_4=-\frac{19}{16}\,\zeta_{4}\,.
\end{equation}

Finally, adding up the results of the various contributions (ladder, ladder-cascade and cascade) to the NGLs coefficient at this order we obtain the result
\begin{equation}
\S^{(4)}_{ab}=-\frac{29}{16}\,\zeta_4\,.
\end{equation}

\section*{Radiator for the hemisphere mass distribution}
\label{app:rad}

The radiator in the $\overline{\mathrm{MS}}$ renormalization scheme for the hemisphere mass distribution is given by (see for instance Ref. \cite {Dasgupta:2002dc})
\begin{align}\label{eq:rad}
\mathcal{R}(\rho)&=\CF\left[L\,r_1(\alpha_sL)+r_2(\alpha_sL)+r_{2,\mathrm{coll}}(\alpha_sL)\right],
\end{align}
with
\begin{subequations}
\begin{align}
r_1&=\frac{1}{2\pi\beta_0\lambda}\left[(1-2\lambda)\ln(1-2\lambda)-2(1-\lambda)\ln(1-\lambda)\right],\\
r_2&=\frac{\mathrm{K}}{4\pi^2\beta_0^2}\left[2\ln(1-\lambda)-\ln(1-2\lambda)\right]+\notag\\
&\quad+\frac{\beta_1}{2\pi\beta_0^3}\left[\frac{1}{2}\ln^2(1-2\lambda)-\ln^2(1-\lambda)+\ln(1-2\lambda)-2\ln(1-\lambda)\right],\\
r_{2,\mathrm{coll}}&=\frac{3}{4}\,\frac{1}{\pi\beta_0}\,\ln(1-\lambda)\,,
\end{align}
\end{subequations}
with $\lambda=\as(Q)\,\beta_0\,\ln(1/\rho)$. Hard-collinear emissions to the outgoing quark initiating the measured hemisphere are accounted for by the term $r_{2,\mathrm{coll}}$ in the radiator. In the expression of the radiator we have the following constants
\begin{equation}
\begin{split}
\mathrm{K}&= \CA\left(\frac{67}{18}-\frac{\pi^2}{6}\right)-\frac{5}{9}\,\mathrm{n_f}\,,\\
\beta_0&=\frac{11\,\mathrm{C_A}-2\,\mathrm{n_f}}{12\,\pi}\,,\\
\beta_1&=\frac{17\,\mathrm{C_A^2}-5\,\mathrm{C_A}\,\mathrm{n_f}-3\,\mathrm{C_F}\,\mathrm{n_f}}{24\pi^2}\,,
\end{split}
\end{equation}
where $\mathrm{n_f}=5$ is the number of active quark flavours. The $L$-derivative of the radiator is given by
\begin{equation}
\mathcal{R}'=\frac{\partial\mathcal{R}}{\partial L}=\frac{\CF}{\pi\beta_0}\left[\ln(1-\lambda)-\ln(1-2\lambda)\right].
\end{equation}

%\nocite{*}
\bibliographystyle{pepan}
\bibliography{BMS}

\end{document}